\documentclass[letterpaper,twocolumn,10pt]{article}
\usepackage{usenix2019_v3}

\usepackage{tikz}
\usepackage{cite}
\usepackage{amsmath,amssymb,amsfonts}
\usepackage{algorithmic}
\usepackage{graphicx}
\usepackage{textcomp}
\usepackage{xcolor}
\usepackage{hyperref}
\usepackage{breakurl}
\usepackage{cleveref}
\usepackage{verbatim}
\usepackage{bytefield}
\usepackage{listings}
\usepackage{tcolorbox}
\usepackage{caption}
\usepackage{subcaption}
\usepackage{dblfloatfix}
\usepackage{flushend}
\usepackage{booktabs} 
\usepackage{url}
\usepackage{svg}
\usepackage{xspace}

\definecolor{dkgreen}{rgb}{0,0.6,0}
\definecolor{gray}{rgb}{0.5,0.5,0.5}
\definecolor{mauve}{rgb}{0.58,0,0.82}
\newcommand*\circled[1]{\tikz[baseline=(char.base)]{
		\node[shape=circle,fill,color=black,text=white,inner sep=0.05pt](char){#1};}}

\makeatletter
\newcommand{\myLabeledItem}[1][]{
        \protected@edef\@currentlabel{#1}%
\item[#1]
}
\makeatother
\newcommand{\accl}{ACCL+\xspace} %use Anon for submission, ACCL later
\newcommand{\acclurl}{https://github.com/Xilinx/ACCL} %Anonimyze for review, replace with https://github.com/Xilinx/ACCL later
\begin{document}
%-------------------------------------------------------------------------------

%don't want date printed
\date{}

\title{\Large \bf \accl: an FPGA-Based Collective Engine for Distributed Applications}

% %for single author (just remove % characters)
\author{
{\rm Zhenhao He}\\
Systems Group, ETH Zurich
\and
{\rm Dario Korolija}\\
Systems Group, ETH Zurich
\and
{\rm Yu Zhu}\\
Systems Group, ETH Zurich
\and
{\rm Benjamin Ramhorst}\\
Systems Group, ETH Zurich
\and
{\rm Tristan Laan}\thanks{Work done during internship at Research Labs AMD Xilinx}\\
University of Amsterdam
\and
{\rm Lucian Petrica}\\
Research Labs AMD Xilinx
\and
{\rm Michaela Blott}\\
Research Labs AMD Xilinx
\and
{\rm Gustavo Alonso}\\
Systems Group, ETH Zurich
}

\maketitle

\vspace{-1ex}
%-------------------------------------------------------------------------------
\begin{abstract}
%-------------------------------------------------------------------------------
FPGAs are increasingly prevalent in cloud deployments, serving as Smart NICs or network-attached accelerators. Despite their potential, developing distributed FPGA-accelerated applications remains cumbersome due to the lack of appropriate infrastructure and communication abstractions.
To facilitate the development of distributed applications with FPGAs, in this paper we propose \accl, an open-source versatile FPGA-based collective communication library. Portable across different platforms and supporting UDP, TCP, as well as RDMA, \accl 
empowers FPGA applications to initiate direct FPGA-to-FPGA collective communication. Additionally, it can serve as a collective offload engine for CPU applications, freeing the CPU from networking tasks. It is user-extensible, allowing new collectives to be implemented and deployed without having to re-synthesize the FPGA circuit. 
We evaluated \accl on an FPGA cluster with 100 Gb/s networking, comparing its performance against software MPI over RDMA. The results demonstrate \accl's significant advantages for FPGA-based distributed applications and highly competitive performance for CPU applications.
We showcase \accl's dual role with two use cases: seamlessly integrating as a collective offload engine to distribute CPU-based vector-matrix multiplication, and serving as a crucial and efficient component in designing fully FPGA-based distributed deep-learning recommendation inference.
\end{abstract}
\vspace{-1ex}
%-------------------------------------------------------------------------------
\section{Introduction}
%-------------------------------------------------------------------------------

FPGAs are increasingly being deployed in data centers~\cite{catapult,catapult2} as Smart NICs~\cite{panic, Corundum,gpu-fpga-nic,diad,nica-fpga-nic}, streaming processors~\cite{stream_processor,Accorda, liu2023honeycomb, STYX}, and disaggregated accelerators~\cite{disaggregated-cloud, KV-Direct, korolija2021farview,faas-fpga,Clio-disagg-mem-fpga,disagg-mem-kona,SmartDS}. In scenarios where FPGAs are directly connected to the network, an opportunity is created for distributed computing with direct FPGA-to-FPGA communication.
However, designing distributed applications with FPGAs is difficult as it requires not only a network stack on the FPGA compatible with the data center infrastructure, but often also higher level abstraction, e.g., \textit{collective communication}, for more complex communication patterns. Unlike in the software ecosystem where many open-source collective libraries exist~\cite{openmpi, MVAPICH}, there is a lack of such resources for FPGA developers. While new development platforms~\cite{kathail2020xilinx, intel_one_api} are improving FPGA programmability, and other recent FPGA virtualization platforms~\cite{Optimus, AmorphOS,FPGA_cloud,fpgavirt,opencl-fpga,vitual-fpga-runtime} focus on virtualizing FPGA resources for abstracting data movement, they lack support for networking. This forces distributed applications on FPGAs to rely on the CPU for networking~\cite{DNNs_on_Multi_FPGA,cnn-aws,cpu-mpi-for-fpga}, significantly increasing the latency of data transfers between FPGAs. It has not been until recently that hardware networking support~\cite{Limago,isca_tcp,strom,easynet,40g_tcp} has emerged in the FPGA ecosystem. However, they all fall short in providing abstractions for collective communication, limiting their applicability in larger distributed use cases. 

\begin{figure}[t]
  \centering
  \includegraphics[width=.45\textwidth]{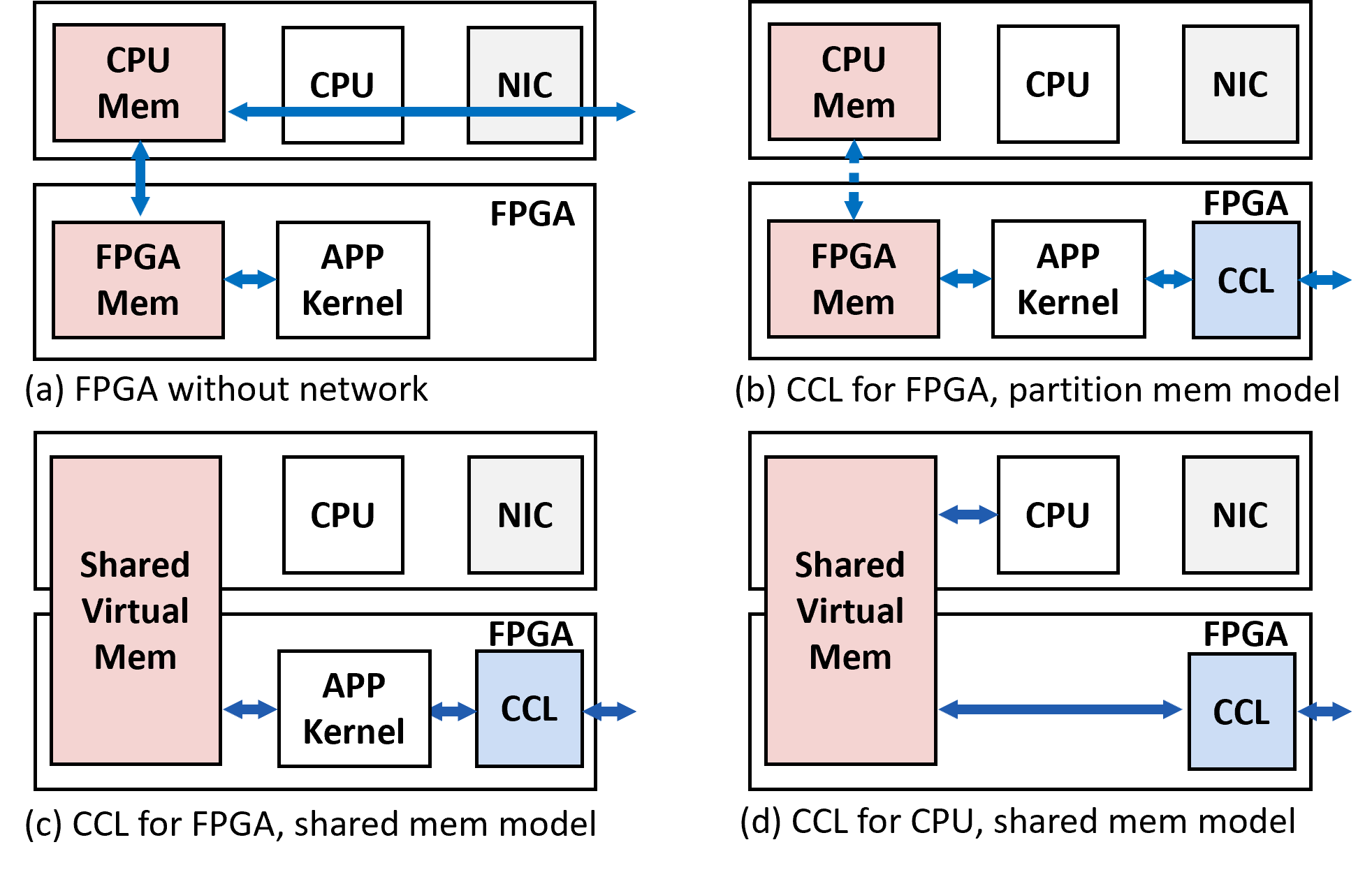}
  % \caption{FPGA-based collective library (CCL) configurations: (a) FPGA application relies on CPU network stack for inter-FPGA data movement. (b) CCL enables direct network for FPGA application with partitioned memory model, involving memory copy between host and device memory (dashed line). (c) CCL for FPGA application with shared virtual memory allows direct access to FPGA/CPU memory space. (d) CCL as an offload engine for CPU applications with shared virtual memory.}
  %\caption{(a) FPGA app relies on CPU network stack for inter-FPGA data movement. (b) FPGA collective communication library (CCL) enables direct networking between FPGAs. Explicit memory copy needed if data origins from CPU memory (dashed line). (c) CCL for FPGA app with shared virtual memory. (d) CCL as an offload engine for CPU app with shared virtual memory.}
  \vspace{-2ex}
  \caption{CCL in different FPGA-accelerated systems.}
  \vspace{-3ex}
  \label{fig:ccl_dataflow}
\end{figure}

Implementing high-performance and versatile collective abstractions for FPGAs poses challenges specific to reconfigurable computing. Portability stands out as a key challenge, as FPGAs are used in quite different configurations. The first aspect of the portability challenge is the \textit{memory model}. FPGAs support various memory models, such as partitioned memory model~\cite{kathail2020xilinx, AmorphOS, intel_quartus} and shared virtual memory model~\cite{coyote, Optimus, pointer-chasing-shared-memory-fpga,harp}, with each memory model being suitable for different scenarios. Figure~\ref{fig:ccl_dataflow} illustrates how an FPGA-based collective communication library (CCL) can be effectively utilized with different memory models. In Figure~\ref{fig:ccl_dataflow}(a), the FPGA application relies on the CPU network stack for inter-FPGA data movement, while Figure~\ref{fig:ccl_dataflow}(b) showcases the FPGA CCL enabling direct networking between FPGAs. If the data originates from CPU memory (dashed line), an explicit memory copy is required for partitioned memory platform. Figure~\ref{fig:ccl_dataflow}(c) illustrates CCL implementation for an FPGA application with shared virtual memory, while Figure~\ref{fig:ccl_dataflow}(d) exemplifies CCL's role as an offload engine for a CPU application with shared virtual memory.

The second aspect is the portability across \textit{communication models} as FPGAs support two fundamentally different ones: message passing, i.e., MPI, where communication occurs between buffers in memory, and streaming, where communication occurs through continuous data streams. The choice of communication model significantly influences how an application interacts with the communication layer, determining whether data needs to be buffered in memory before communication~\cite{ACCL,zrlmpi,MPI_transpilation}, or if it can be streamed directly to the communication layer~\cite{smi,LEAP,PASTA}. Existing FPGA communication frameworks tend to address only one of them.

The third aspect of the portability involves supporting diverse communication protocols due to application-specific needs and ensuring interoperability in heterogeneous environments where FPGAs coexist with CPUs and accelerators. This flexibility ensures seamless integration and communication between FPGA-based offloaded components and other system elements. Existing works~\cite{hw_mpi_barrier,smi,walters2007mpi,zrlmpi,MPI_transpilation,matrix-mul-collective-fpga,OmpSs, OPTWEB, ViTAL,fpga-Hyperscale} that provide collective abstractions on FPGAs or provide multi-FPGA support are often tailored for scenarios where FPGAs are directly connected to each other rather than integrated into a data center packet-switched network. These approaches communicate through low-level link-layer protocols, leading to scalability and integration challenges at a data-center scale. 

In summary, the key question to address is \textit{how to effectively design a portable, flexible, high-level collective abstraction on FPGAs that can support various memory models, communication models, and transport protocols, while accommodating a broad spectrum of applications}. Achieving this objective is complex, given the significant impact of these configurations on runtime, interfaces, and data movement. Moreover, given FPGAs' extended compilation times and lengthy hardware debugging cycles, we need a parameterized approach that allows a FPGA-resident CCL to be modified without recompilation, in order for the CCL to be practical in real-world use.
% Despite the challenges, such an approach is also critical for expanding the collective abstraction across the FPGA boundary, eventually to CPU and GPU clusters, as indicated by recent work on partitioning workload between FPGAs, CPUs and GPUs~\cite{fleetrec, gpu-fpga-nic}. 

To address these challenges, we introduce \accl, a hardware collective communication library (CCL) for FPGAs. \accl provides MPI-like collective APIs % to applications, offering  collective APIs for the message-passing paradigm 
with explicit buffer allocation and streaming collective APIs with direct channels to the communication layer. The modular system architecture decouples platform-specific IO management and runtime from the collective implementation, incorporating platform and network protocol-specific adapters and drivers. Additionally, we have developed a platform and protocol-independent collective offload engine that supports  modifying the collective implementation without hardware recompilation. \accl can be used to 
%can serve as a collective abstraction for FPGA applications, 
enable direct communication between FPGAs and can function as a collective offload engine for the CPU. We test \accl on two platforms, a commodity, partitioned memory platform (AMD Vitis~\cite{kathail2020xilinx}) and a shared virtual memory platform (Coyote~\cite{coyote}). 
% \accl support common communication protocols, including UDP, TCP and RDMA.
We choose AMD Vitis for its recent integration of high-performance 100 Gb/s UDP~\cite{vnx} and TCP~\cite{easynet} hardware stacks, aligning with our goal of leveraging cutting-edge networking capabilities for optimal communication performance. Coyote is used due to its unique provisioning of unified and virtualized memory across CPU-FPGA boundaries~\cite{coyote}, coupled with comprehensive network services, including RDMA. 
\accl is also designed to minimize control overheads, improve scalability, and facilitate simulation.

We first evaluate \accl using micro benchmarks. \accl achieves a peak send/recv throughput of 95 Gbps, almost saturating the 100 Gb/s network bandwidth. We evaluate collective operations under two scenarios: FPGA-to-FPGA distributed applications with FPGA kernels directly interacting with \accl (F2F), and CPU-to-CPU distributed applications with \accl as a collective offload engine (H2H). \accl exhibits significantly lower latency in F2F scenarios compared to software RDMA MPI for FPGA-generated data. In H2H scenarios, \accl has comparable performance to software RDMA MPI for CPU-generated data while freeing up CPU cycles and reducing pressure on CPU caches. Then, we examine \accl with two use-case scenarios. First, distributed vector-matrix multiplication with CPU computation and \accl-based reduction, where \accl improves performance compared to software MPI. Secondly, we show that \accl enables the distribution of an industrial recommendation model across a cluster of 10 FPGAs, achieving more than two orders of magnitude lower inference latency and more than an order of magnitude higher throughput than CPU solutions. The use case study not only highlights \accl's effectiveness in different scenarios but also paves the way for future research opportunities in investigating hybrid CPU-FPGA co-design for distributed applications.

\vspace{-1ex}
\section{Background}\label{sec:background}

\noindent\textbf{FPGA Programming.}
In the past, hardware description language (HDL - Verilog, VHDL) was the sole method to program FPGAs. With the emergence of High-Level-Synthesis (HLS), the programmability of FPGAs is enhanced by allowing the developers to program in C-like code with hints (pragmas) to infer parallel hardware blocks. 
Despite the advancements, existing HLS-based libraries lack networking and collective abstractions.

\noindent\textbf{Communication Models.} Message passing, represented by MPI, is a well-established communication model in distributed programming on CPUs, whereby communicating agents exchange messages, i.e., user buffers, typically resulting from previous computation. This model can be applied to FPGAs, but a more common communication model for FPGAs is the streaming model. FPGA kernels support direct streaming interfaces, into which data can be pushed in a pipelined fashion during processing. Kernels executing on the same FPGA can stream data to each-other through low-level latency-insensitive channels, such as AXI-Stream~\cite{axisspec}. The streaming model can be applied for communication across FPGAs, however, existing streaming communication framework~\cite{LEAP,latency-insensitive-channel,smi} often do not have transport protocols or collective abstractions.

\noindent\textbf{FPGA Development Platforms.}
Modern FPGA platforms adopt various virtualization methodologies~\cite{fpga-virtualization-survey, fpga-virtualization-survey-2,fpga-cloud-survey} for FPGA resources. Most simplify development with a static \emph{shell} for resource management and data movement, with some offering additional \textit{services} like transport layer networking, and the host-device interaction relies on \textit{runtime} libraries. This approach allows developers to concentrate on designing the application kernel. Many commodity platforms~\cite{kathail2020xilinx, intel_quartus} implement a partitioned memory model, which permits data movement from FPGA applications to FPGA memory while restricting direct access to host CPU memory. In contrast, shared virtual memory platforms, such as Coyote~\cite{coyote} and Optimus~\cite{Optimus}, offer an abstraction of virtualized and unified memory space across CPU and FPGA that the FPGA kernel can directly access. This also facilitates RDMA to CPU memory from FPGA network streams.

\noindent\textbf{Network-Attached FPGAs.}
The data center FPGA landscape has evolved with FPGAs featuring 100 Gb/s transceivers, enabling direct processing of network data~\cite{catapult, net_fgpa_dc}. FPGA-based Smart NICs~\cite{Corundum, panic,isca_tcp, gpu-fpga-nic,lynx, sume, wave-runner,electronics9010059,FlexDriver} perform programmable packet filtering but often leave the network stack to the CPU software, limiting their applicability to FPGA applications. Distributed machine learning~\cite{elastic-df, BrainWave, fpl_recommendation_cluster,multi-fpga-nn,scale-out-ml-fpga}, and data processing~\cite{KV-Direct,fpga-net-intrusion-detection, korolija2021farview} applications capitalize on network-attached FPGAs and thepir communication pattern tends to be more complicated with larger designs, motivating the need for high-level collective abstractions on FPGAs. Following this trend, there is an increasing effort to develop hardware network stacks on FPGAs, such as UDP~\cite{vnx,fpga-udp}, TCP~\cite{10g-tcp,sidler2016lowlatencytcp,40g_tcp, Limago}, and RDMA~\cite{fpga-rdma,strom,multi-path-fpga-rdma,KV-Direct}. However, only network stack offload is often not enough for complicated applications in distributed settings. 
\vspace{-1ex}
\section{Related Work}\label{sec:related_work}

% \noindent\textbf{Message Passing for FPGAs.}
MPI implementation of collective communication has evolved to become more accelerator-aware over time, e.g., GPU-aware MPI~\cite{gpu_aware_mpi,yang2011hybrid,potluri2012optimizing}. Some GPU-direct MPI implementations enable direct data transfers between GPU memories through RDMA NIC, bypassing host memory~\cite{NCCL,RCCL}. However, FPGA-based collective is different as FPGAs can connect directly to the network, removing the need for an external NIC. Additionally, standard remote DMA with commodity NICs lacks support for streaming interfaces to FPGA kernels, forcing streaming kernels to buffer data before collective operations. Projects such as Galapagos~\cite{Galapagos,heterogeneous_cloud} and EasyNet~\cite{easynet} provide in-FPGA communication stacks for data exchange within a cluster, serving as a foundation for collectives without an external NIC. TMD-MPI~\cite{TMD-MPI,tmd-mpi-ecosystem} orchestrates in-FPGA collectives using embedded processors, yet its bottleneck lies in control due to sequential execution in low-frequency FPGA microprocessors. Collective offload with NetFPGA~\cite{collective_netfpga,Arap2014SoftwareDM,zynq_collective} has been explored, but static collective offload engines limit flexibility and often rely on software-defined network switches for orchestration.
SMI~\cite{smi} proposes a streaming message-passing model, exposing streaming collective interfaces to FPGA kernels. While SMI enables kernels to initiate collectives directly, it employs dedicated FPGA logic for collective control, limiting flexibility for post-synthesis reconfiguration. ACCL~\cite{ACCL} focuses primarily on message-passing collectives for FPGA applications. The coordination of collectives requires CPU involvement, lacking significant streaming support, and has not been tested at scale. BluesMPI~\cite{blues_mpi, bluesmpi+} offloads collective operations to a BlueField Smart-NIC, demonstrating comparable communication latency to host-based collectives, but it does not target accelerator applications. The latency of \accl targeting host data matches BluesMPI, even with BluesMPI ARM cores working at ten times the frequency.

% \noindent\textbf{Multi-FPGA Frameworks.}
Frameworks like ViTAL and its successors~\cite{ViTAL, hetero-vital, multi-layer-vital} propose FPGA resource virtualization and compilation flows for mapping large designs onto multiple FPGAs through latency-insensitive channels. OmpSs@cloudFPGA~\cite{OmpSs} introduces a multi-FPGA programming framework that partitions large OpenMP programs with domain-specific programs into smaller distributed parts for execution on FPGA clusters, providing communication through static, compile-time-defined send/recv and collective operations supporting only the unreliable UDP protocol. Elastic-DF~\cite{elastic-df} and FCsN~\cite{FCsN} present domain-specific frameworks for automatically distributing large neural network model inference across FPGAs with hardware UDP/TCP send/recv for FPGA-to-FPGA data movement. These projects are complementary to our work, and integrating \accl as their FPGA communication backend can enhance their flexibility and performance.
\vspace{-1ex}

\section{\accl: An FPGA Collective Engine}

\begin{figure}
  \centering
  \includegraphics[width=.4\textwidth]{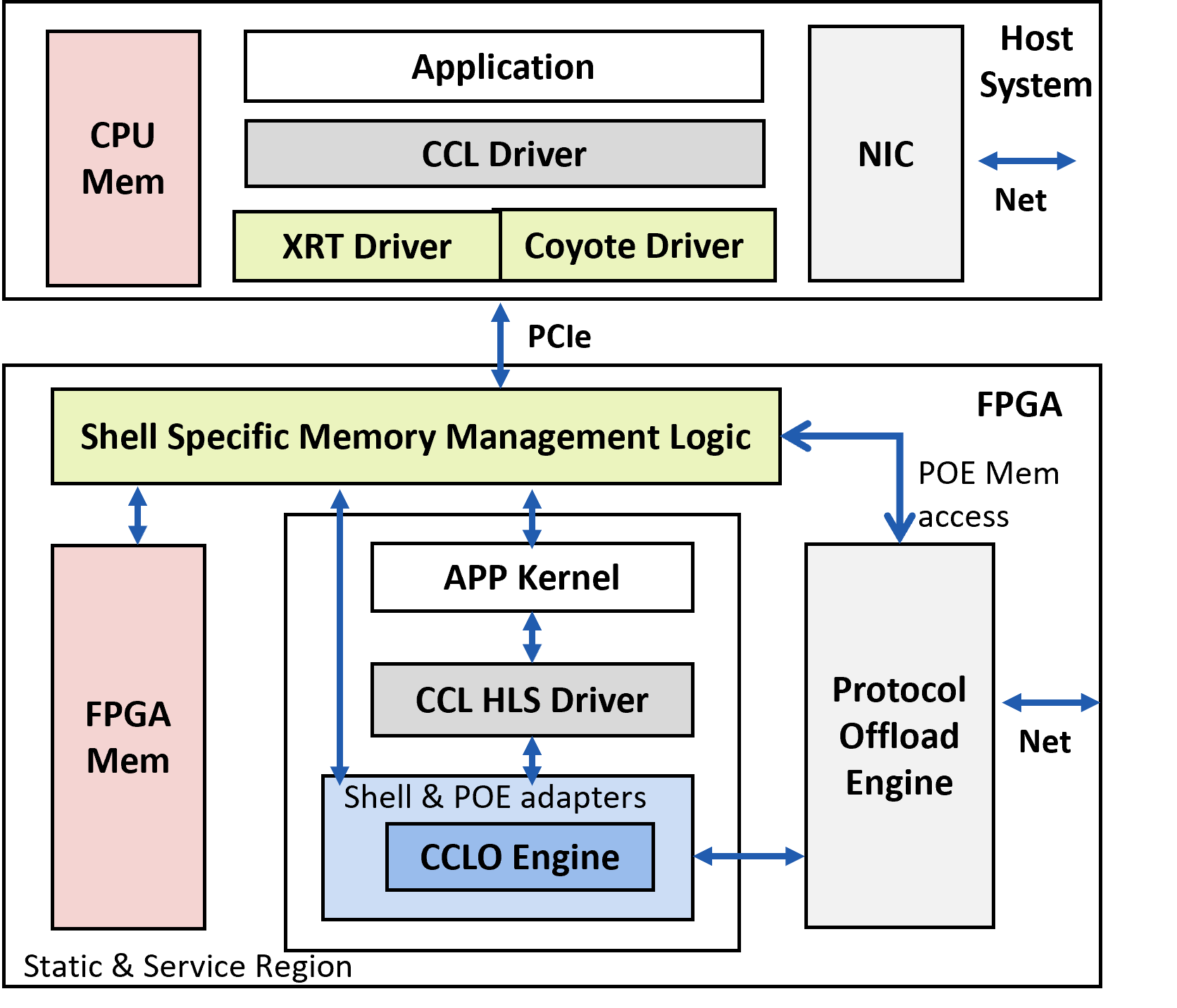}
  \caption{System overview of the FPGA-based collective communication library.}
  \vspace{-3ex}
  \label{fig:system_overview}
\end{figure}
The architecture of \accl\footnote{\acclurl} is modular, featuring layers of abstraction in both software and hardware, as illustrated in Figure~\ref{fig:system_overview}, facilitating adaptability of a central \accl CCL offload engine (CCLO) to diverse platforms and communication protocols. In this section we will describe each layer.

\subsection{Application Interface}

\noindent\textbf{\accl Drivers.}
FPGA-accelerated applications execute partially on the CPU and partially on FPGA, whereby each of the CPU- and FPGA-resident components of the application may require access to collective communication infrastructure. For this reason \accl implements two drivers which expose similar platform- and protocol-agnostic collective APIs to each of these two components. The host-side \accl CCL driver allows initialization and runtime management of platform and \accl data structures and hardware, as well as POE initialization, i.e., setting up sessions or queue-pairs. The CCL HLS driver is not capable of such initialization, and therefore the application must perform host-side initialization before any FPGA application kernels are started.

%\subsection{MPI-like and Streaming Collective APIs}
% - API provided as HLS library 
% - APIs that we support currently 
% - easy to extend or add a new API -> corresponds to the modification within the micro-processor firmware
% - Listings of APIs
\noindent\textbf{MPI-like Collective API.} 
\accl drivers expose an MPI-like API, catering to the message-passing paradigm and facilitating the porting of existing MPI-based applications to \accl collectives. This API require the application to store data in memory before invoking collectives. Listing~\ref{lst:reduce} shows the MPI-like collective API, including arguments like datatype, buffer pointer, and element count, along with flags indicating buffer location (host or FPGA memory) and the option for synchronous calls. To facilitate portability, message passing collectives operate on an \accl specific buffer class which can wrap normal C++ arrays with additional platform-specific information. Common collectives, such as reduce, broadcast, and barrier, are supported. Each MPI-like collective call in the host CCL driver has a corresponding HLS API call with a similar syntax for direct invocation from FPGA kernels.

\begin{lstlisting}[label={lst:reduce},caption={Reduce collective API in C++.}]
CCLRequest *reduce(dataType src_data_type, BaseBuffer &buf, unsigned int count, unsigned int root, reduceFunction func, communicatorId comm_id, flagType flags);
\end{lstlisting}

\noindent\textbf{Streaming Collective API.}
\accl also provides a streaming collectives API whereby data originates and terminates at the stream interfaces between the FPGA application kernels and the \accl hardware component, instead of memory buffers. The HLS-based streaming APIs are tailored for FPGAs applications running in streaming fashion.

\begin{lstlisting}[label={lst:stream_send},caption={Example kernel using streaming send in HLS.}]
//set up command and data interfaces
cclo_hls::Command cclo(cmd, sts, communicator);
cclo_hls::Data data(data_to_cclo, data_from_cclo);
//issue streaming send command without buffer argument
cclo.send(type, count, dst_rank);
// push data in streams to network without buffering
for (int i = 0; i < N; i++) {
    data.push(/* generate data */); }
cclo.finalize(); // wait for send completion
\end{lstlisting}

Listing~\ref{lst:stream_send} demonstrates an example FPGA kernel issuing a streaming send command to the CCLO engine (line 5) and subsequent pushes to the CCLO streaming data interface, 64B per cycle (line 8), followed by a wait for CCLO completion. This code is synthesizable with HLS tools.
HDL-based FPGA kernels can
interact with the collective engine directly, through the same interfaces. Additionally, the host can also call streaming collectives via the host-side CCL driver.

\subsection{\accl Platform Support}

\begin{figure}
  \centering
  \includegraphics[width=.45\textwidth]{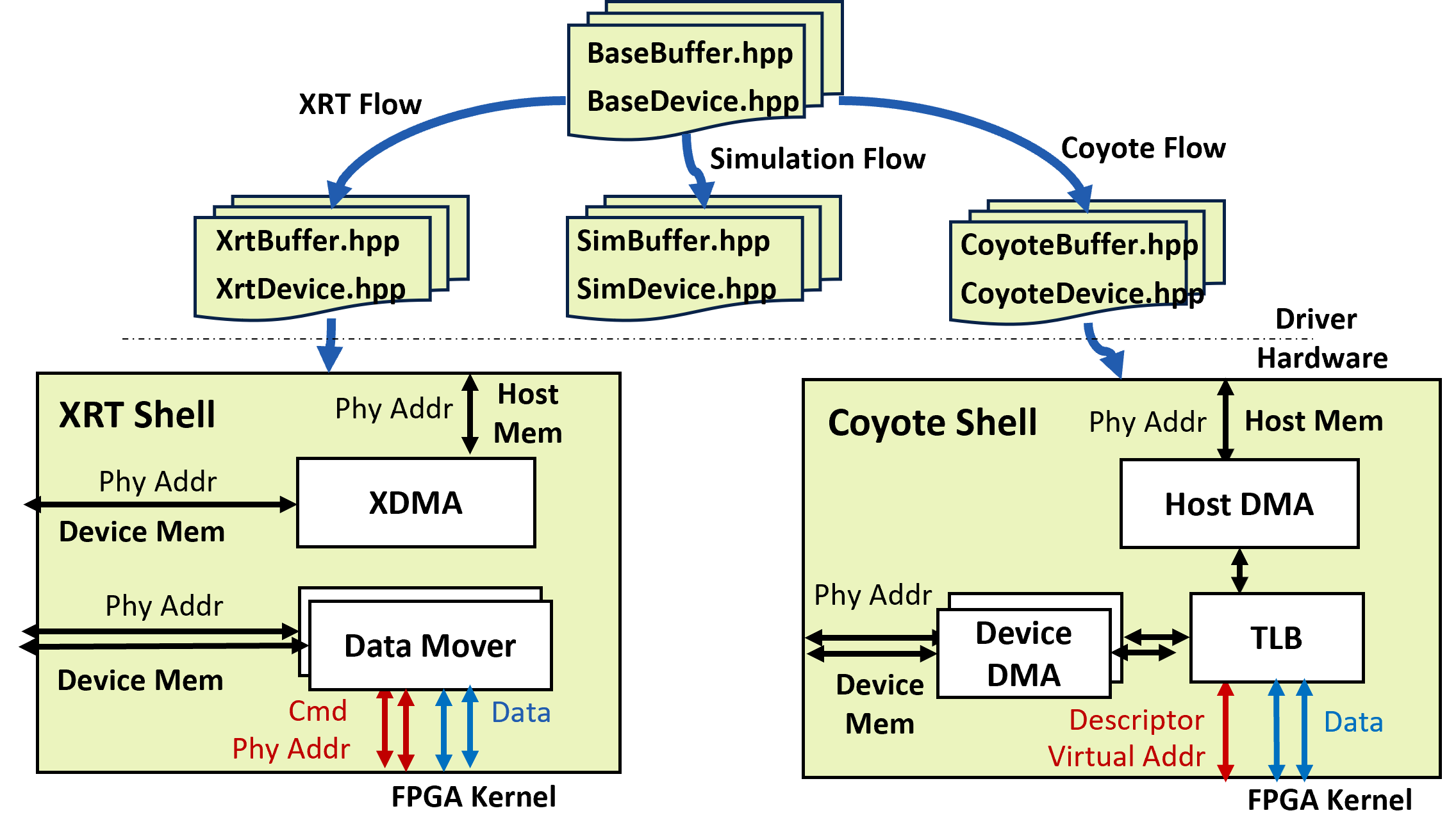}
  \caption{CCL driver for different memory managements.}
  \vspace{-2ex}
  \label{fig:driver_shell}
\end{figure}

A platform is defined by a software interface specification, defining how FPGA memory is allocated and manipulated, and how FPGA kernels are called, and a hardware interface specification, i.e., how FPGA kernels, including the CCLO, plug into hardware services in the FPGA.
To facilitate portability between platforms, the \accl host CCL driver layers the APIs on top of generic class types, such as BaseBuffer for memory allocation and data movement between host and FPGA, and BaseDevice for CCLO invocation. These are specialized to individual platforms through class inheritance, as illustrated in Figure~\ref{fig:driver_shell}. Each specific CCL class interfaces with platform-native drivers and employs distinct processes for handling data movement.
\accl supports both the commercial AMD Vitis platforms and the open-source Coyote platform~\cite{coyote}, as well as a virtual simulation platform. New platforms can be added easily.

\noindent\textbf{Integration with Coyote.} Coyote utilizes a shared-memory model with a central memory management logic governed by a software-populated translation lookaside buffer (TLB). This TLB records allocated pages and facilitates virtual-physical address translation. The FPGA kernel issues memory requests through a descriptor interface, using virtual addresses directed to either host or device memory. The TLB interprets these requests, interacting with host DMA or device DMA based on the physical location of the memory page, forwarding the data to FPGA applications in a streaming manner. If a memory page is not registered during TLB lookup, it triggers an interruption to the CPU, resulting in a page fault and introducing a performance penalty. Therefore, the CCL driver, specifically the CoyoteBuffer class, eagerly maps pages to the Coyote TLBs when instantiating buffers. We modified Coyote, during integration, to increase the associativity of the TLB cache and expand the number of streaming interfaces Coyote provided to a single application region, from a single interface to three interfaces which is required by the CCLO engine. We also implemented a Coyote-specific adapter to convert from CCLO (R)DMA request syntax to Coyote-specific syntax, as indicated in Figure~\ref{fig:rdma_adapter}.

\noindent\textbf{Integration with Vitis.} Vitis platforms implement a partitioned memory model and the Xilinx Runtime (XRT)~\cite{kathail2020xilinx} is utilized by the \accl CCL driver for low-level interaction with the platform. A XRT-controlled XDMA IP core~\cite{xdma} moves data between host and FPGA memory, while FPGA memory is accessed by FPGA kernels through Data Movers~\cite{datamover}. We implemented dedicated adapters which include Data Movers for compatibility of the CCLO Engine with Vitis platform memory interfaces. 
As a result of the partitioned memory, the CCL driver explicitly migrates buffers between host and FPGA memory prior to or after the collective execution if the data originally resides in host memory - a process denoted staging. Staging creates performance penalty when \accl collectives target host memory, as observed by related work on collective offload on DPUs~\cite{bluesmpi+}. Therefore, Vitis platforms favor distributed FPGA applications where data is streamed or resides in FPGA memory.

\noindent\textbf{Simulation Platform}
We implemented an additional simulation platform for debugging and performance optimization. This simulation platform roughly models a Vitis platform, whereby FPGA chip interfaces (XDMA, Ethernet) are replaced by ZMQ~\cite{hintjens2013zeromq} interfaces. A stand-alone simulated FPGA node is compiled to include memory and one \accl CCLO Engine. The \accl host driver includes dedicated buffer and device abstractions capable of connecting to the simulated node via ZMQ. \accl provides convenient launch scripts to set up a simulated cluster of such simulation nodes.  

The simulated nodes connect to each other through ZMQ rather than real Ethernet. While the simulated ZMQ network may lack realistic features like packet loss and reordering, it serves as a valuable functional simulation.

\accl provides two simulation levels of the CCLO engine: functional simulation using compiled \accl HLS source code and C firmware, and cycle-accurate (but slow) simulation using Verilog HDL generated from compiling the CCLO HLS code and firmware. 
For FPGA applications requiring streaming data exchange between FPGA kernels and the CCLO, we provide a bus functional model of the CCLO that connects via ZMQ to the simulated node.

\subsection{Protocol Offload Engine}

\begin{figure}[t]
  \vspace{-5pt}
  \centering
  \includegraphics[width=0.4\textwidth]{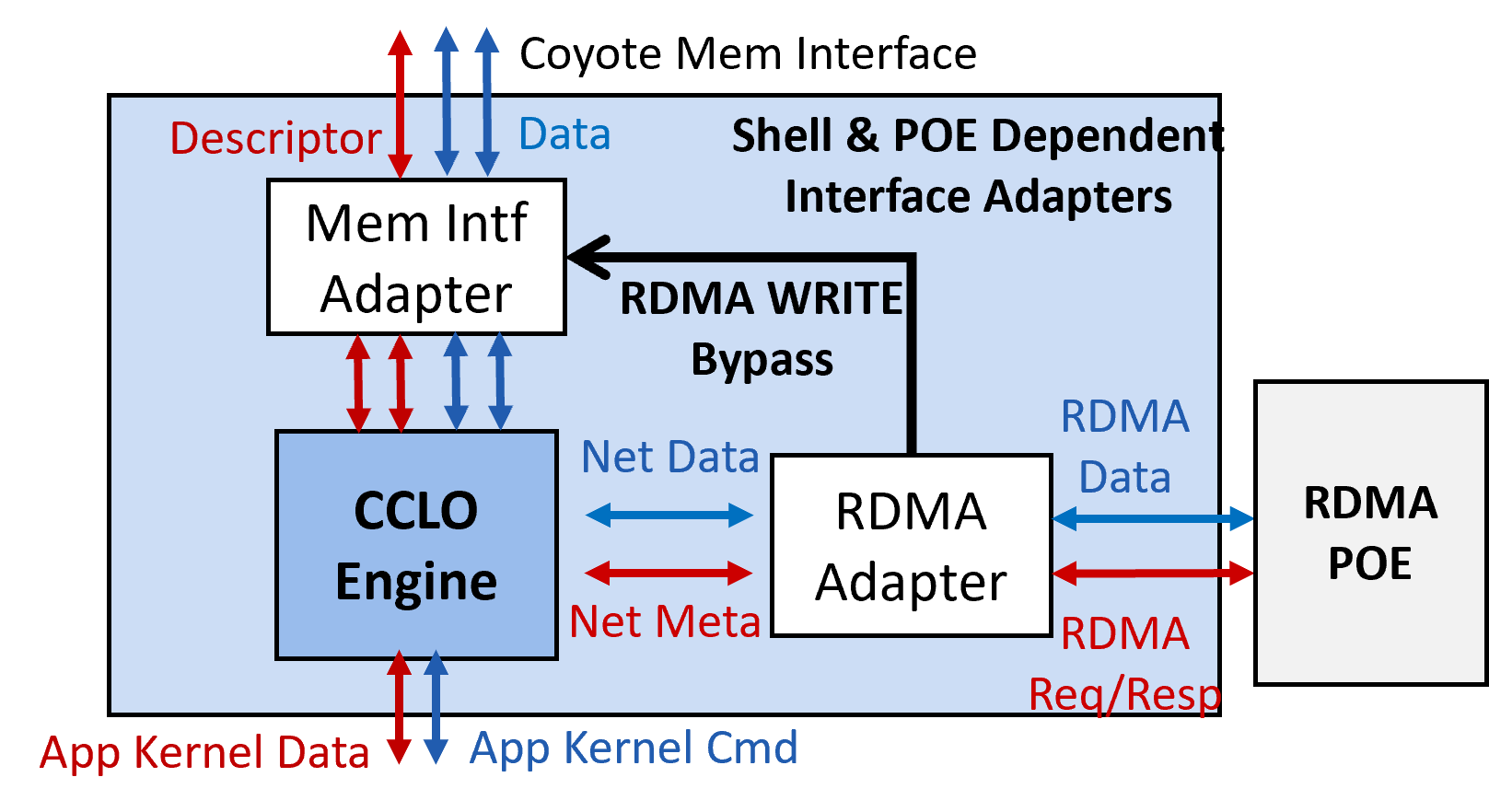}
  \caption{\accl with Coyote-RDMA data path with corresponding POE and memory adapters.}
  \vspace{-1.0\baselineskip} 
  \label{fig:rdma_adapter}
\end{figure}

\accl supports several 100 Gb/s protocol offload engines (POE) in hardware: UDP~\cite{vnx} and TCP~\cite{easynet} on Vitis platforms, and all the network services provided by Coyote. Notably, \accl supports collectives with RDMA by leveraging the unified and virtualized memory space across the FPGA and the CPU provided by Coyote. All the POEs expose streaming control and data interfaces to other modules and some POEs (e.g., TCP) require direct memory access for packet buffering for re-transmission. For portability, the CCLO Engine has a set of POE-independent internal interfaces - two pairs of meta and data streaming interfaces (one for Tx and one for Rx). The meta interfaces contains various sub fields to indicate the op code, data length, communication session IDs, etc. The meta interfaces are then adapted to the POE interfaces with dedicated FPGA components as exemplified in Figure~\ref{fig:rdma_adapter}. The selection of the POE and its adapters is a compile time parameter of the CCLO Engine. 

\noindent\textbf{Coyote RDMA POE.} It supports standard RDMA verbs, including one-sided operations like \texttt{WRITE} and two-sided operations like \texttt{SEND}. The RDMA POE incorporates various streaming interfaces for RDMA commands, memory commands, and data.
\noindent\textit{Default Configuration:} On the initiating side of a \texttt{WRITE} operation, the RDMA POE issues memory requests directly to the Coyote memory management logic, fetching data from either host or device memory and streaming it through the network. On the passive side of \texttt{WRITE}, the data is directly written to virtualized memory.
\noindent\textit{\accl Integration:} In the \accl-enabled configuration, the CCLO engine acts as a "bump-in-the-wire" engine between the memory management unit and the RDMA POE, as shown in Figure~\ref{fig:rdma_adapter}. On the initiating side of a \texttt{WRITE}, the CCLO engine issues RDMA commands and is responsible for data preparation, either fetching from memory or obtaining it from the application kernel in the form of streams. On the passive side, data bypasses the CCLO and is directly forwarded to the memory management unit. For single-sided \texttt{WRITE}, streaming into the application kernel is also possible by configuring the datapath at compile time. The CCLO engine consistently manages data and metadata streams from two-sided \texttt{SEND}. For CCL driver with RDMA, a queue pair needs to be exchanged between each node and needs to be registered to the RDMA POE.

\noindent\textbf{TCP POE.} The TCP POE supports up to 1,000 connections and can be configured to support window scaling and out-of-order packet processing. As a reliable transmission protocol, the TCP POE also needs to access protocol-internal buffers for re-transmission. The CCLO engine prepares and accepts all the data streams with the TCP POE. For CCL driver with TCP POE, a TCP session needs to be established between each node to construct the communicator.

\subsection{CCLO Engine}

\begin{figure}[h]
  \vspace{-5pt}
  \centering
  \includegraphics[width=0.4\textwidth]{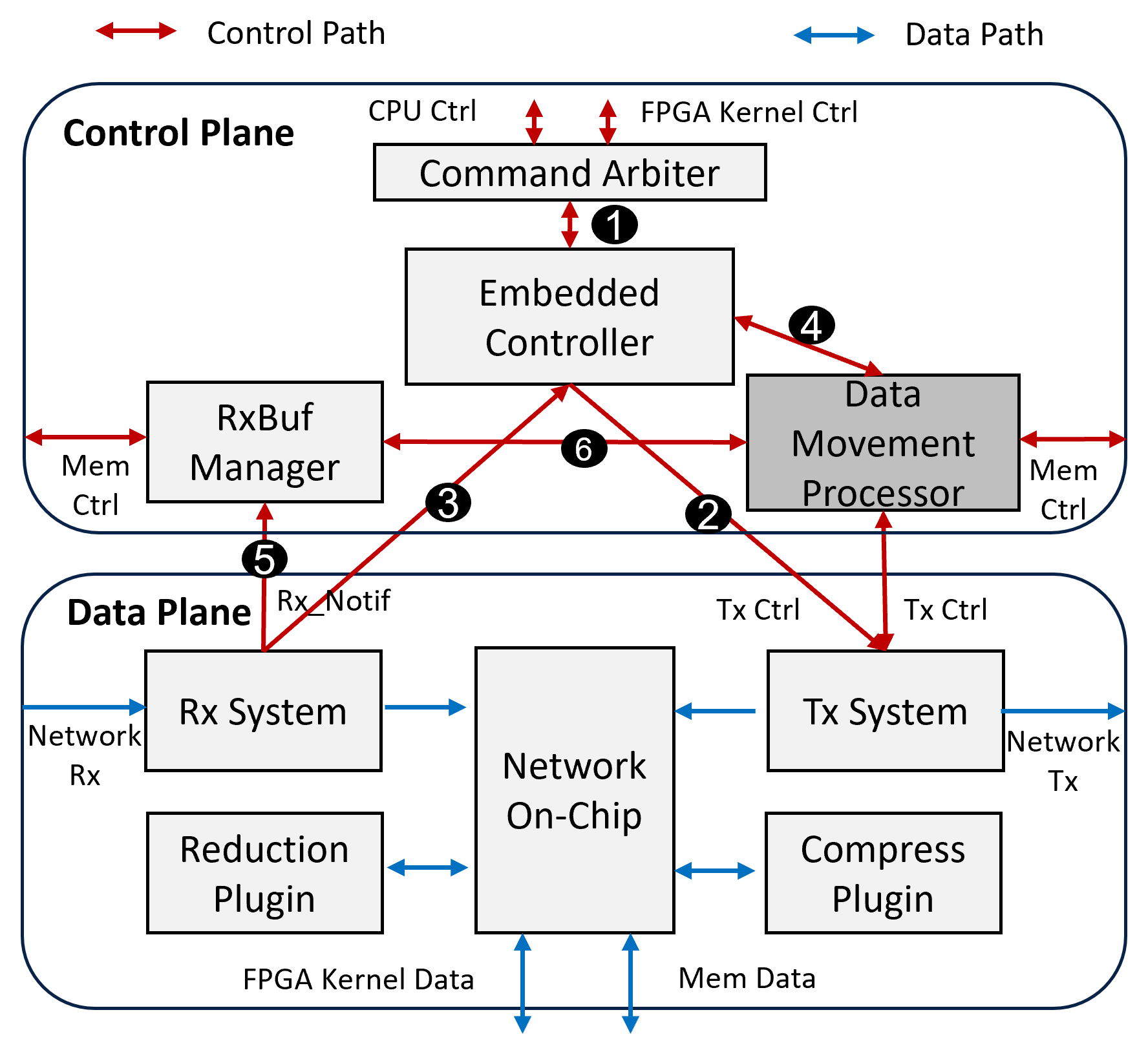}
  \caption{Hardware architecture of CCLO engine, where \protect\circled{1} is collective commands, \protect\circled{2} and \protect\circled{3} are commands for rendezvous protocol initialization and done, \protect\circled{4} stands for microcode for data movement, \protect\circled{5} and \protect\circled{6} are commands to manage Rx buffers for eager protocol.}
  \vspace{-1.0\baselineskip} 
  \label{fig:cclo_overview}
\end{figure}

The CCLO Engine orchestrates the collective data movement through a set of standardized CCLO interfaces. The CCLO accepts communication requests from the host or application kernels, communicates with the protocol offload engine, manages buffers in FPGA memory (HBM, DDR, BRAM), and manage data streams from other kernels. 
To enable simultaneous flexibility, low latency, and high throughput, the key design principle is to \textit{decouple the CCLO logic into the flexible control plane and the parallel data processing plane}, as indicated in Figure \ref{fig:cclo_overview}. 

\subsubsection{Flexible Control Plane}\label{control_plane_description}

% - Center part is the micro-processor
% - Microprocessor memory implemented as BRAM
% - communicator offload with CPU
% - collective implemented with control sequences
% - control sequences
% - example collective implementation with the control sequence and algorithm

The CCLO control plane is flexible, centered around an embedded micro-controller (uC)~\cite{microbalze}, which responds to commands from other FPGA kernels or the host and issues high-level data movement commands to a hardware-accelerated data movement processor (DMP) and other components. The CCLO control plane also contains a RxBuf Manager (RBM), which manages temporary Rx buffers. The uC, DMP, and RBM store states in a small configuration memory implemented as FPGA BRAM. The configuration memory is also accessible by the CPU through MMIO and includes information about the communicator, e.g., session or queue pair IDs, pool of allocated Rx buffers.

\noindent\textbf{Embedded Controller.} The uC firmware implements different collective algorithms and different synchronization protocols, e.g., eager and rendezvous. Though all collectives could also be implemented in digital logic, the uC provides the high flexibility to implement different collective algorithms by updating the firmware without the need to refactorize the whole design and re-synthesize. Considering the sequential execution nature of the uC and a low clock frequency, to avoid too much overhead in the uC, the firmware is carefully designed such that it only issues a set of coarse-grained control commands to other hardware control blocks, which are highly parallel and latency-optimized. Besides, FIFO queues are incorporated into all command paths, allowing multiple in-flight instructions.

\noindent\textbf{Data Movement Processor.} The DMP accepts and executes microcode generated by the uC \circled{4} according to collective algorithms. Each microcode instruction has three slots: two operand slots and one result slot. Each microcode slot describes a data movement behavior to be executed by the data plane. Operand slots describe data coming into the CCLO, while the result slot describes data movement out of the CCLO. If the operand is expected to come over network and buffered in temporary buffers, the DMP also sends out requests periodically to the RBM to check if the message has arrived \circled{6}. For example, a send operation uses an operand slot for data movement from FPGA memory into the CCLO and a result slot for data leaving the CCLO into the network POE. Similarly, reduction involves all three DMP instruction slots for orchestrating data movement through two input buffers, arithmetic plugin, and finally to the result buffer.
The DMP operates in a pipelined fashion, and each operand slot independently interprets its instruction fields, emitting commands for corresponding datapath blocks. Upon receiving acknowledgements from datapath blocks, the DMP signals instruction completion to the uC.

\noindent\textbf{RxBuf Manager.} 
The RBM is used when temporary buffering is needed, e.g., collective with eager protocol. Upon the notifications of incoming messages from the network, as shown in path \circled{5} in Figure~\ref{fig:cclo_overview}, RBM retrieves a list of available Rx buffers from the configuration memory and then it issues memory requests to store the message in the selected Rx buffer. As messages are typically divided into network packets, and packets from different sessions may arrive in an interleaved manner, the RBM also contains data structures to reassemble the packets into messages in corresponding Rx buffer. The RBM also stores relevant metadata (source ID, tag, Rx buffer address) to be used by the DMP. 

\subsubsection{Parallel Data Plane}\label{data_plane_description}

\noindent\textbf{Rx and Tx System.}
In \accl, we implement a lightweight communication protocol above the transport layer protocol to carry metadata information for each message. Each message consists of a signature and a payload consisting of user data. The signature contains the rank IDs of the message, message type, source and destination, message length, tag, a sequence number which is used to keep track of the order of the messages and other meta information. The Tx and Rx systems are responsible for packetizing and depacketizing the signature along with user payload, and they issue commands to interact with the POEs. The command issuing, signature insertion, and parsing processes can vary for different synchronization protocols. Both the Rx and Tx systems incorporate a finite state machine to respond appropriately to these variations.

\noindent\textbf{Network On Chip.} All the data streams internal to the CCLO can be routed in packets based on the dest field that comes along with the data.  

\noindent\textbf{Streaming Plugins.} The plug-ins are utilized for applying unary and binary operations to in-flight data and can be enabled at compile time. Binary operations are typically utilized to implement reductions - sum, max, etc. Unary operators may implement compression or encryption. Each of the plug-ins is a streaming kernel and may implement more than one function, in which case the control plane will specify the desired function by setting its dest field of the plugin input stream.

\subsubsection{Message Synchronization Protocol} \label{section:synchronization}

\begin{figure}[t]
	\centering
	\vspace{-1.5ex}
	\subfloat[Eager protocol]{\includegraphics[width=1.63in]{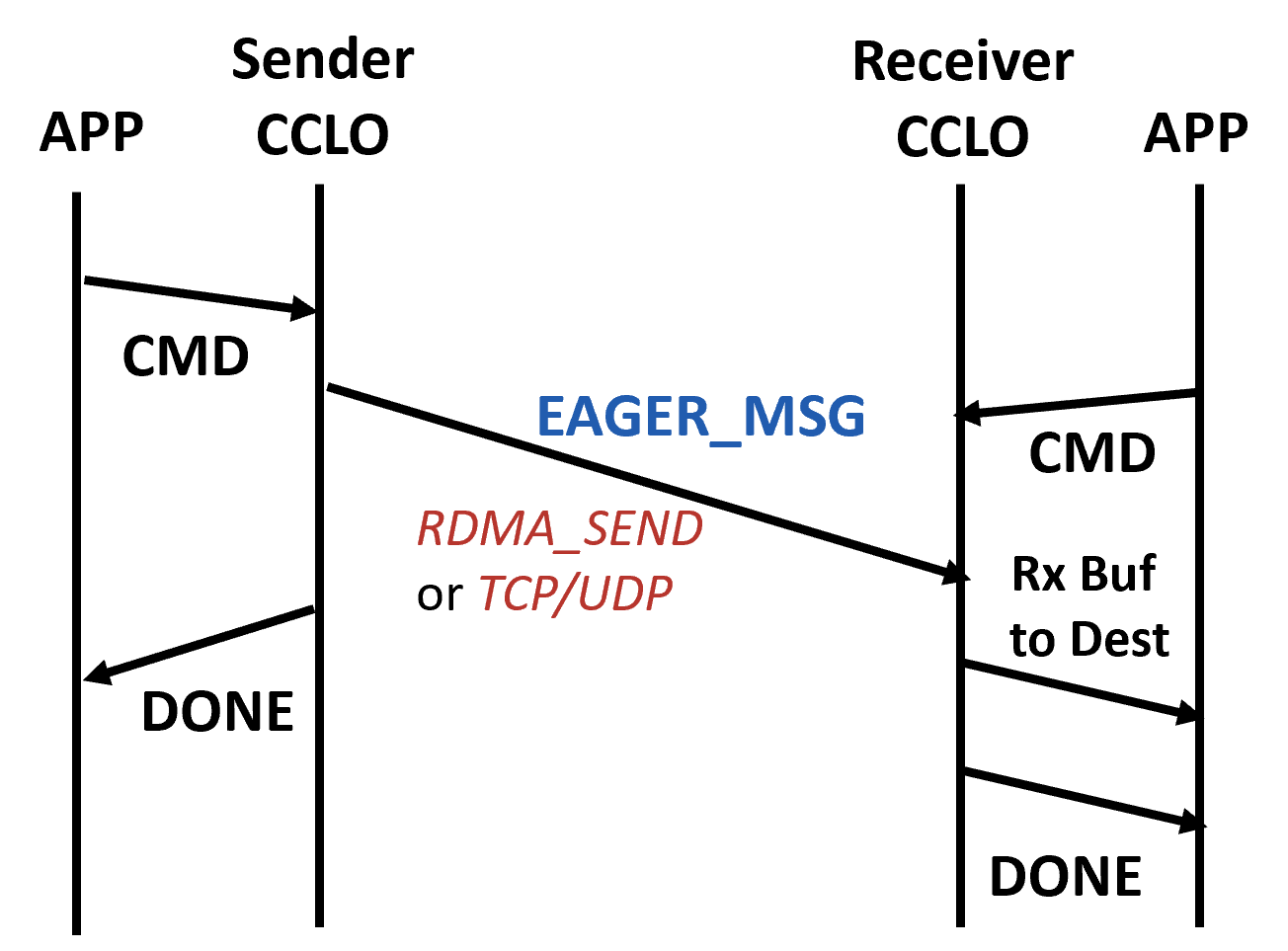} 
		\label{eager}} 
	\subfloat[Rendezvous protocol]{\includegraphics[width=1.63in]{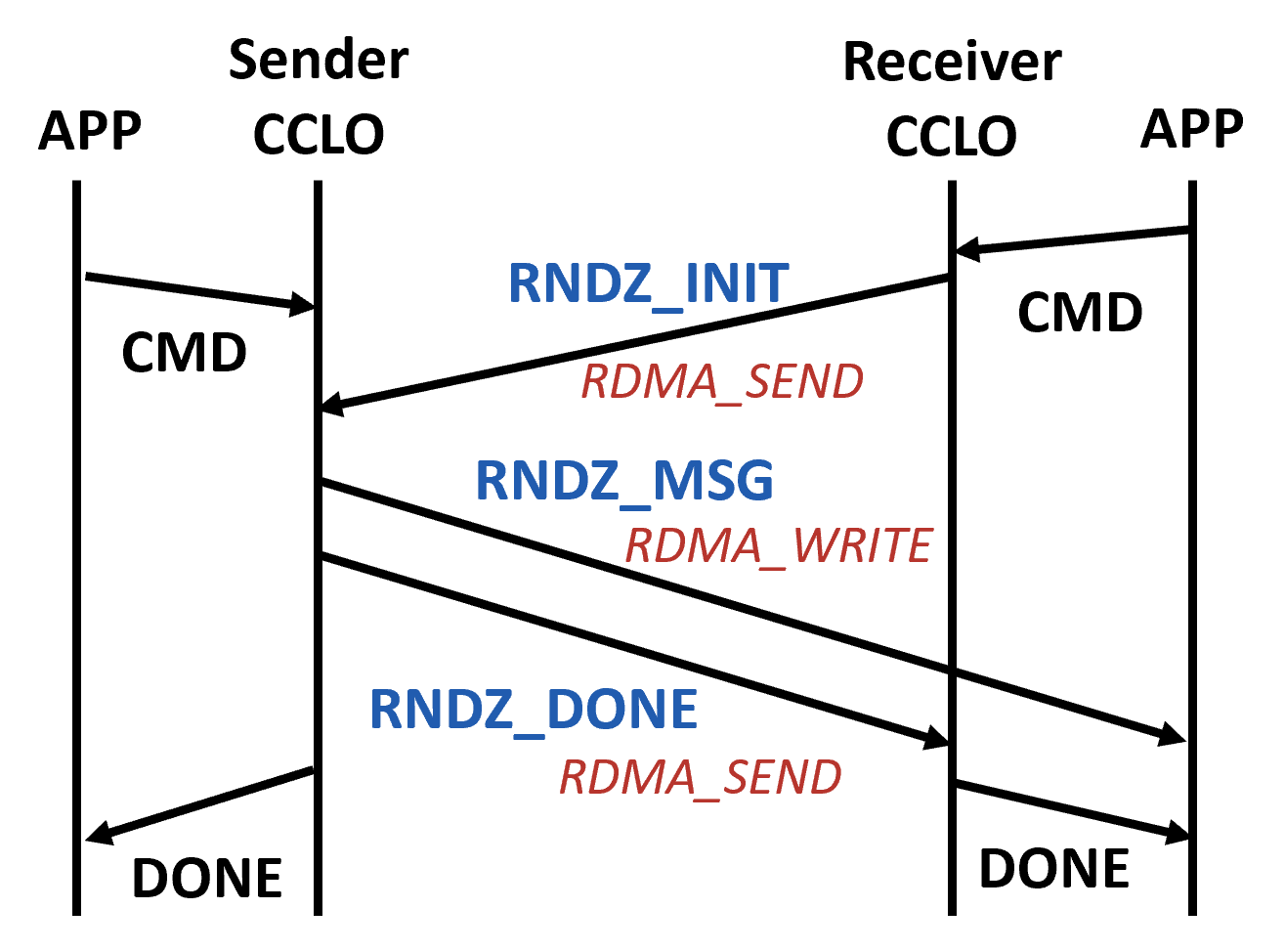} 
		\label{rndz}} 
		\vspace{-1.5ex}
		\caption{CCLO eager and rendezvous with send/recv.}
		\label{syn_protocol} 
	\vspace{-3ex}
\end{figure} 

Eager protocol allows sender CCLO to send as soon as the command arrives and the data at the receiver is buffered in the CCLO Rx buffer before it is moved to the destination (either in memory or in FPGA kernel streams depends on runtime configuration), as shown in Figure~\ref{syn_protocol}(a). Eager protocol is sometimes favored, especially with small messages, as the lack of a handshake phase minimizes latency as copies incur little overhead at small message sizes. We implement eager protocol with UDP/TCP or two-sided RDMA \texttt{SEND} verb. In contrast, rendezvous protocol first requires the result buffer address resolution before the actual message transmission, as shown in Figure~\ref{syn_protocol}(b). Once the destination buffer is resolved, the data can be directly put into the destination buffer, avoiding the cost of temporary buffering at Rx Buffer. We use two-sided RDMA \texttt{SEND} verb for rendezvous handshake messages, such as \texttt{RNDZ\_INIT} and \texttt{RDNZ\_DONE}, and we use RDMA \texttt{WRITE} for actual message transmission.

% \subsubsection{Execution Walk Through} \label{section:execution}
Here we use examples to illustrate the control and data sequence for different protocols. 
\noindent\textbf{RDMA Rendezvous Send/Recv:} 
Upon the receipt of invocation command on the Recv rank, the uC initiates a Tx control command containing the result buffer address to the Tx system \circled{2}. The Tx system generates an \texttt{RDZV\_INIT} message to the Send rank. The Rx system of the Send rank queues up a notification with the result address for the uC \circled{3} to pick up. The uC of Send rank issues a microcode with \texttt{RNDZ\_MSG} type to the DMP, which fetches data from memory and issues a Tx control to the Tx system. The Tx System issues an RDMA \texttt{WRITE} command to the POE, forwarding the data to the network. Once the RDMA \texttt{WRITE} is complete, the Tx System issues an \texttt{RDZV\_DONE} message with RDMA \texttt{SEND}. On the receiving side, the \texttt{RNDZ\_MSG} is directed straight to the memory, bypassing the receiver CCLO intervention. The subsequent \texttt{RDZV\_DONE} is forwarded to the Rx system and then to the uC \circled{3}.
\noindent\textbf{TCP/RDMA Eager Send/Recv:}
Upon the command's arrival at the uC, it issues a microcode to the DMP \circled{4}, which issues memory commands and a Tx control with type \texttt{EAGER\_MSG}. The Tx system waits for the data from the memory and appends a header with the message signature. On the receiving side, the Rx System receives the eager message, parses the header, and forwards it to the RBM \circled{5}. The RBM accesses the exchange memory, which stores the RxBuf list. If there is an empty Rx buffer, it issues memory commands to buffer the input message. When the receive command arrives at the uC, it issues microcode to the DMP \circled{4}, which checks the RBM for the expected message \circled{6}. If the message is already stored in an Rx buffer, the DMP initiates a memory control operation to transfer the data to the result buffer.

\subsubsection{Collective Algorithms}
We provide different implementations for various collectives, and users can define their own. Collectives are realized by specifying a communication pattern as a C function in uC firmware, and then executing this pattern through instructions in DMP and Tx/Rx System on each FPGA in the communicator. Table~\ref{table:collective-algo} summarizes the algorithms and communication patterns being used to implement collectives. For eager protocols with unreliable transmission (e.g., UDP), we currently use simple algorithms like ring and one-to-all to minimize the chances of packet loss. Future firmware improvements can enhance POE awareness for finer-grained algorithm selection.
In contrast, when using RDMA, the rendezvous protocol employs more advanced algorithms like tree or recursive doubling. The token-based flow control in RDMA makes it well-suited for these sophisticated algorithms in the rendezvous protocol. For broadcast, we implement a simple one-to-all algorithm with small rank size, while with large rank size, we adopt more advanced recursive doubling such that the data transmission is not bottlenecked at the root rank. For gather and reduce, we apply a similar strategy. With small message size, we adopt an all-to-one approach to reduce the number of intermediate hops needed. On the other hand, with larger message sizes, to avoid a potential in-cast problem with the all-to-one approach, we adopt a tree-based algorithm. The tuning of the algorithms for specific collective can be done at runtime by setting configuration parameters to the CCLO engine and we set these parameters according to our empirical experiment results.

\begin{table}[htbp]
\vspace{-2ex}
\centering
\caption{Algorithms used for example collectives.}
\vspace{-1ex}
\label{table:collective-algo}
\begin{tabular}{|c||c|c|}
 Collective & Eager & Rendezvous\\ 
 \hline
 Bcast &One-to-all&One-to-all;Recursive doubling\\ 
 \hline 
 Reduce & Ring & All-to-one;Binary tree \\    
 \hline
 Gather & Ring & All-to-one;Binary tree \\   
 \hline
 All-to-all & Linear & Linear \\   
\end{tabular}
\vspace{-2ex}
\end{table}

\vspace{-1ex}
\section{Microbenchmark Evaluation}
We evaluate \accl on a heterogeneous cluster with 10 AMD EPYC CPUs and 10 attached FPGA cards (Alveo-U55C). Each CPU is equipped with a 100 Gb/s Mellanox NIC, while each FPGA features a 100 Gb/s Ethernet interface. All devices are connected to Cisco Nexus 9336C-FX2 switches.
Evaluation scenarios consider data residing on the FPGA for distributed FPGA application (suffix \textit{F2F}) and on the CPU for distributed CPU applications (suffix \textit{H2H}). For F2F, the FPGA application data traverses the network directly through \accl. As a baseline, the FPGA data initially is moved to the CPU memory and then is transmitted via a commodity NIC. In H2H, the CPU application data is transferred to the FPGA and then transmitted with \accl. This is compared to transmitting the CPU data directly with a commodity NIC. The focus of these experiments is evaluating RDMA running with Coyote (suffix \textit{cyt}) due to space limitations. We nevertheless present some results with \accl running TCP on top of the XRT platform to compare it to ACCL [16], which utilizes an embedded micro-controller to orchestrate collective operations. Experiments configure both MPI-like collectives with memory pointers and streaming collectives. For the H2H experiments, MPI-like collectives are mandatory, while the F2F experiments are configured to run with streaming collectives. \accl operates at 250 MHz in micro-benchmarks, with varying frequency in the use-case study due to the design complexity. 
% TCP POE has an MTU of 1535 bytes, and RDMA MTU is set at 4096 bytes. 
The comparisons involve MPICH 4.0.2 with TCP and OpenMPI 4.1.3 compiled with RDMA using OpenUCX 1.13.1 on the cluster CPUs and Mellanox 100 Gb/s NICs. MPI libraries self-configure for collective algorithms and synchronization protocols. Each micro benchmark experiment is averaged over 250 runs.

\begin{figure}
  \centering
    \includegraphics[width=.4\textwidth]{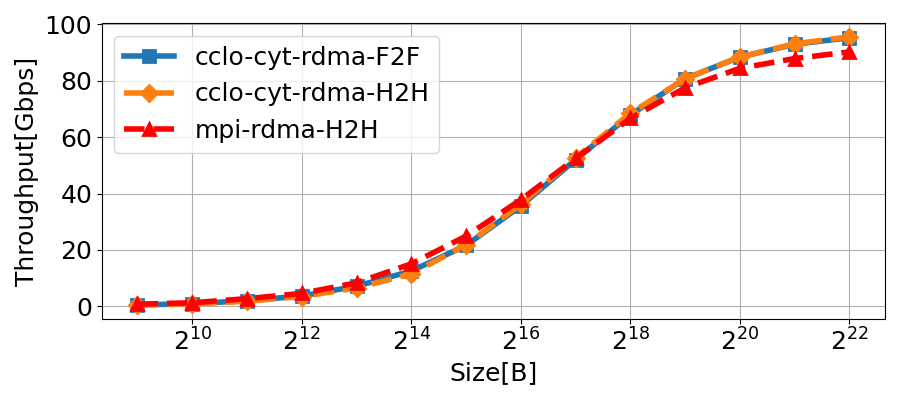} 
    \caption{Send/Recv throughput comparison.}
    \vspace{-2ex}
    \label{fig:sendrecv_bench}
\end{figure}

\begin{figure}
  \centering
    \includegraphics[width=.45\textwidth]{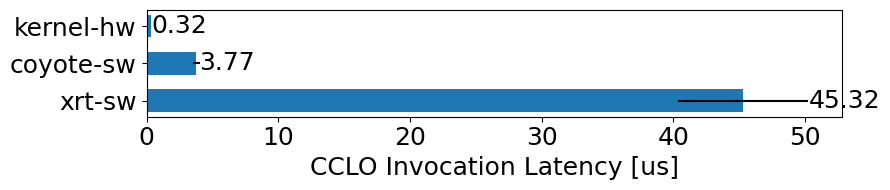} 
    \caption{CCLO invocation latency from different parts.}
    \vspace{-2ex}
    \label{fig:invocation_latency}
\end{figure}

\begin{figure}
  \centering
    \includegraphics[width=.4\textwidth]{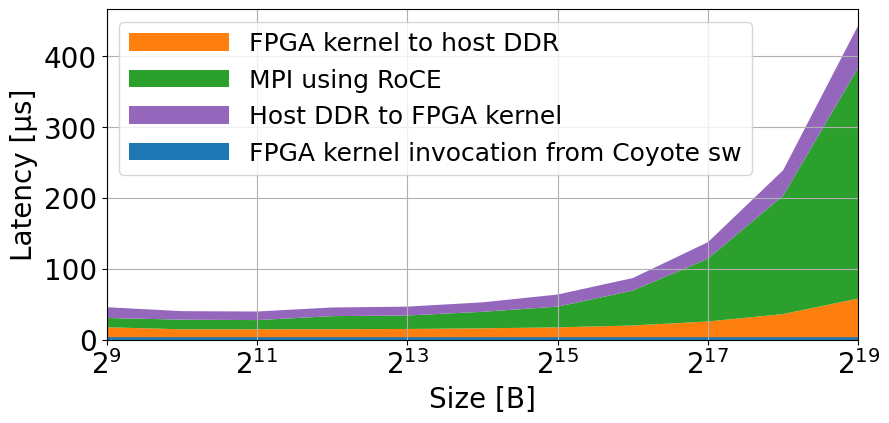} 
    \caption{Latency breakdown of broadcasting FPGA produced data using software MPI with eight ranks with Coyote.}
    \vspace{-2ex}
    \label{fig:latency_breakdown_mpi}
\end{figure}

%\subsection{Micro Benchmark}

\begin{figure*}[t]
     \centering
     \begin{subfigure}[b]{0.245\textwidth}
         \centering
         \includegraphics[width=\textwidth]{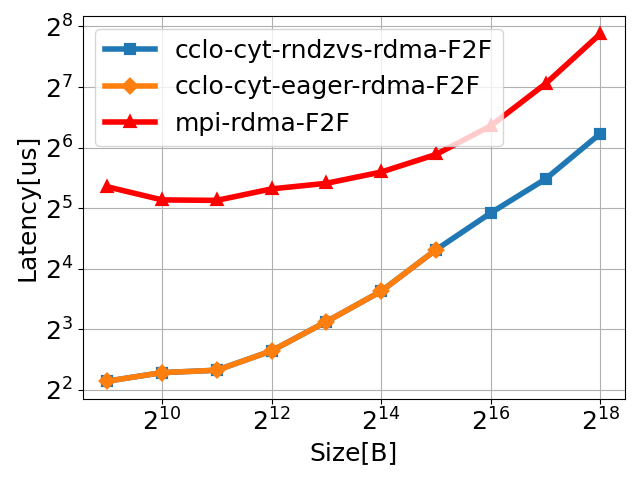}
         \caption{Broadcast}
         \label{fig:bcast_bench_rdma_device}
     \end{subfigure}
     \hfill
     \begin{subfigure}[b]{0.245\textwidth}
         \centering
         \includegraphics[width=\textwidth]{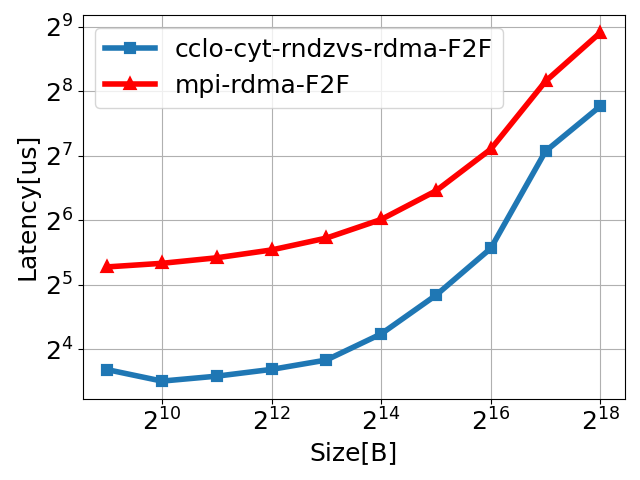}  
         \caption{Gather}
         \label{fig:gather_bench_rdma_device}
     \end{subfigure}
     \hfill
     \begin{subfigure}[b]{0.245\textwidth}
         \centering
         \includegraphics[width=\textwidth]{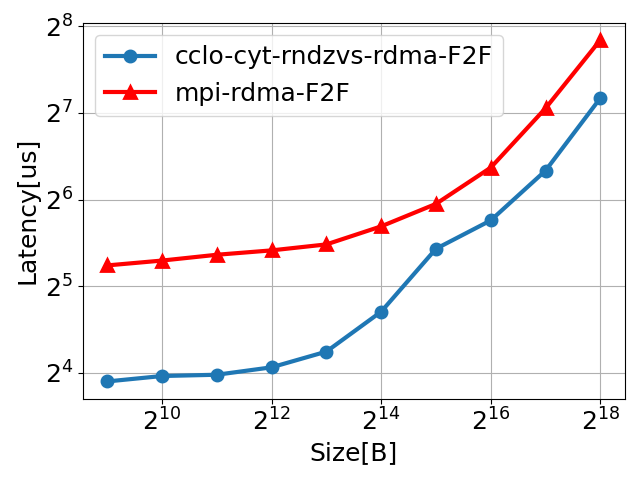}  
         \caption{Reduce}
         \label{fig:reduce_bench_rdma_device}
     \end{subfigure}
     \hfill
     \begin{subfigure}[b]{0.245\textwidth}
         \centering
         \includegraphics[width=\textwidth]{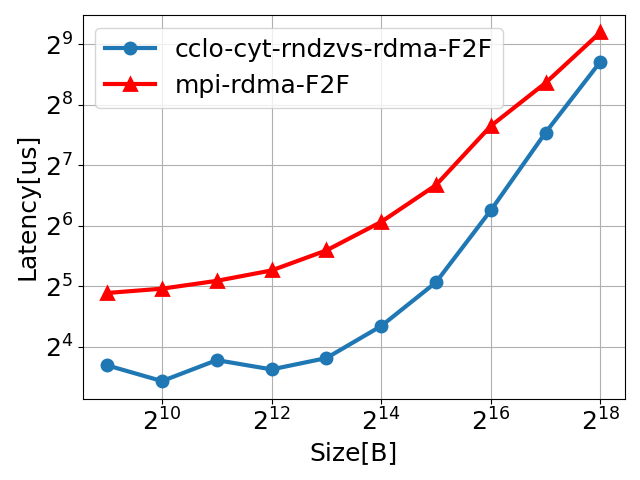} 
         \caption{All-to-all}
         \label{fig:alltoall_bench_rdma_device}
     \end{subfigure}
     \vspace{-1ex}
    \caption{Collective latency comparison between \accl RDMA and software MPI RDMA with eight ranks and device data}
    \vspace{-1ex}
    \label{fig:collectives_bench_rdma_device}
\end{figure*}

\begin{figure*}[t]
     \centering
     \begin{subfigure}[b]{0.245\textwidth}
         \centering
         \includegraphics[width=\textwidth]{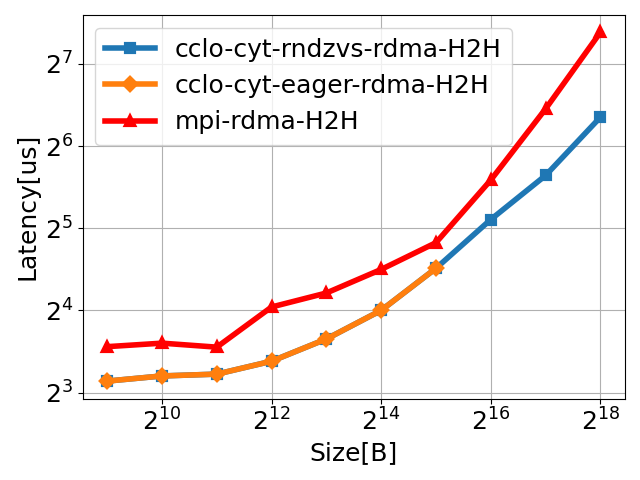}
         \caption{Broadcast}
         \label{fig:bcast_bench_rdma_host}
     \end{subfigure}
     \hfill
     \begin{subfigure}[b]{0.245\textwidth}
         \centering
         \includegraphics[width=\textwidth]{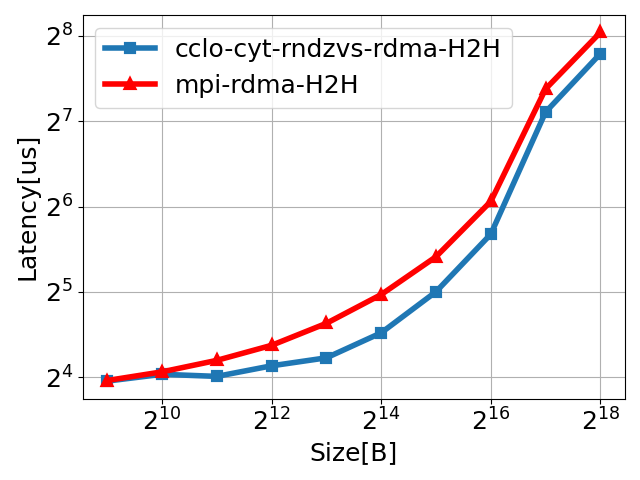}  
         \caption{Gather}
         \label{fig:gather_bench_rdma_host}
     \end{subfigure}
     \hfill
     \begin{subfigure}[b]{0.245\textwidth}
         \centering
         \includegraphics[width=\textwidth]{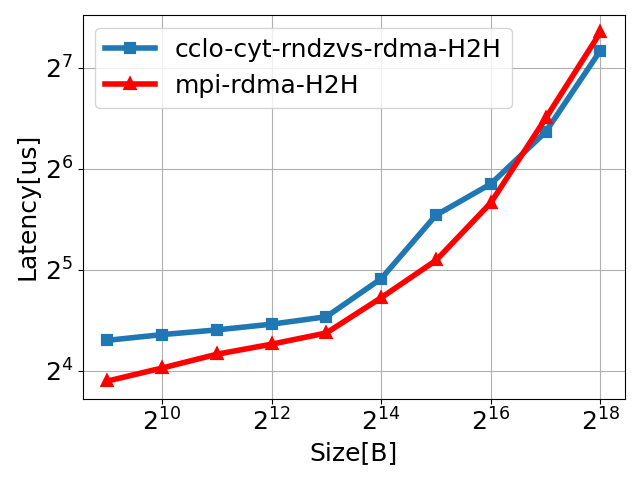}  
         \caption{Reduce}
         \label{fig:reduce_bench_rdma_host}
     \end{subfigure}
     \hfill
     \begin{subfigure}[b]{0.245\textwidth}
         \centering
         \includegraphics[width=\textwidth]{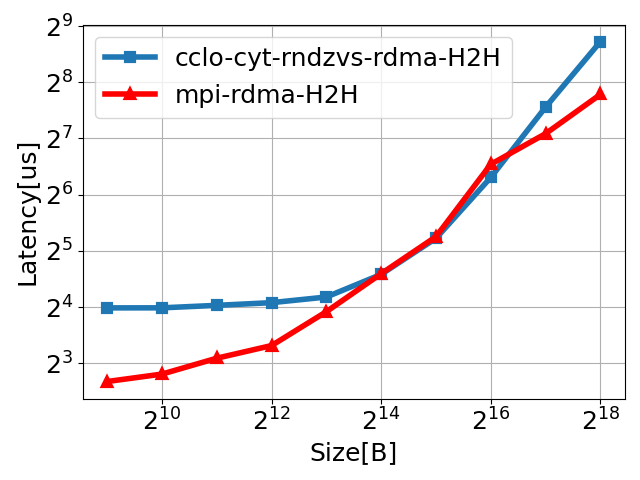}  
         \caption{All-to-all}
         \label{fig:alltoall_bench_rdma_host}
     \end{subfigure}
    \vspace{-1ex}
    \caption{Collective latency comparison between \accl RDMA and software MPI RDMA with eight ranks and host data}
    \vspace{-2ex}
    \label{fig:collectives_bench_rdma_host}
\end{figure*}

\noindent\textbf{Send/Recv Throughput.} We first evaluate pure throughput using send/recv. Figure~\ref{fig:sendrecv_bench} shows the throughput comparison of \accl with Coyote RDMA and software MPI with RoCE backend. Notably, \accl with RDMA achieves a peak throughput of 95 Gb/s, nearly saturating the network bandwidth. Compared to software MPI variants, \accl exhibits comparable and slightly higher peak throughput. This is attributed to the FPGA network stack's ability to process network packets at line-rate in a pipelined fashion. 
Moreover, there is minimal distinction between F2F and H2H for \accl, thanks to the unified memory space provided by Coyote and both host memory access through PCIe and FPGA memory access offer higher bandwidth than the network. 

\noindent\textbf{Invocation Latency.} Figure~\ref{fig:invocation_latency} shows the invocation latency of the CCLO engine to execute a dummy NOP operation, which includes the time from receiving request untill the acknowledgement. For FPGA kernels that can directly interact with the CCLO engine, the invocation latency is minimal compared to software invocation from the host, showing a clear benefit of bypassing host control with FPGA-based applications. Coyote software driver contains a thin and optimized layer for invocation and scheduling and the resulting CCLO invocation time mainly consists of a PCIe write and a PCIe read latency. In contrast, the XRT invocation latency is significantly higher as it is not intended for fine-grained data movement.

\noindent\textbf{FPGA-to-FPGA with software MPI.} To enable a more direct \accl vs. software MPI comparison for executing collectives between kernels on FPGA, we model the execution time for MPICH- and OpenMPI-based device-to-device data movement, which includes: (1) moving data from FPGA HBM/kernel to host DDR through the PCIe, (2) executing the collective using software MPI, (3) moving data from host DDR to FPGA HBM/kernel, and (4) invoking the next computation kernel. We use the CCLO host invocation time as an approximation of the invocation time of other computation kernels. We measure the duration of each of the above steps and present a break-down of execution time of the collective with Coyote platform in Figure~\ref{fig:latency_breakdown_mpi}. We could observe that the PCIe transfer time is dominant for small messages while the collective time is dominant for large messages. Such breakdown for XRT platform can be derived by changing the Coyote invocation latency to XRT invocation latency.

\noindent\textbf{F2F Collective Latency RDMA.} Figure~\ref{fig:collectives_bench_rdma_device} illustrates the latency of \accl RDMA collectives with various message sizes on eight Alveo-U55C boards. This is compared to FPGA-to-FPGA data movement with software MPI over RDMA. For clarity, we present experiments showcasing better performance between eager and rendezvous collectives. The algorithms for each collective in \accl are detailed in Table~\ref{table:collective-algo}. Notably, \accl exhibits significant performance benefits compared to its software counterpart. This advantage stems from the hardware's efficient execution of collectives and the direct network access within the FPGA device, eliminating the need for data copying to CPU memory.

\noindent\textbf{H2H Collective Latency RDMA.} Figure~\ref{fig:collectives_bench_rdma_host} illustrates a latency comparison between \accl and software MPI targeting host data. The performance gains with \accl vary across different collectives. Notably, for operations like broadcast and gather, \accl consistently outperforms software MPI across a range of message sizes. However, for other collectives such as reduce and all-to-all, \accl shows only marginal benefits and, in some cases, falls short of software MPI performance. One reason is that \accl is clocked at much lower frequency compared to commodity-NIC and another reason is that software MPI adapts its algorithms more finely to different configurations, whereas \accl currently supports only a limited set of options. However, by offloading collective to hardware, CPU cycles could be freed for other computation tasks. Besides, by comparing \accl F2F and H2H performance, we could observe that the \accl collective latency has minimal difference because Coyote with unified memory allows direct memory access to both host and FPGA memory. 

\noindent\textbf{Effect of Synchronization Protocol.} Despite the simpler algorithms used by most eager collectives, such as one-to-all or ring, we observe that eager collectives can sometimes outperform rendezvous collectives with small message sizes, as seen in broadcast. This is because eager collectives do not require a handshake to resolve addresses.

\begin{figure}[t]
	\centering
	\subfloat[8KB message size.]{\includegraphics[width=1.625in]{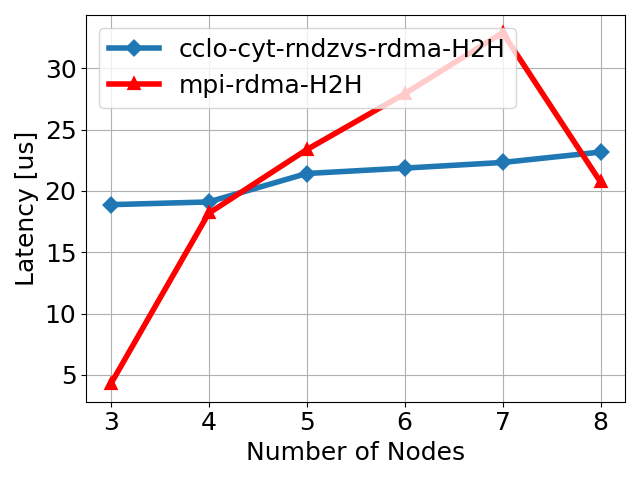} 
		\label{fig:scalability_gather}} 
	\subfloat[128KB message size.]{\includegraphics[width=1.625in]{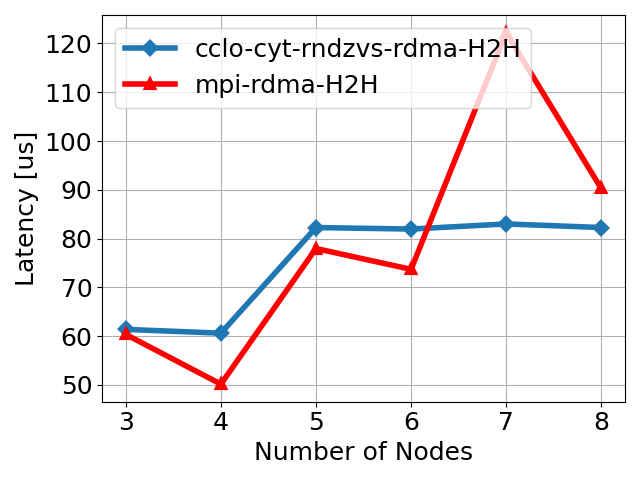} 
		\label{fig:scalability_reduce}} 
		\vspace{-1.5ex}
		\caption{ Latency vs. rank sizes (Reduce).}
            \label{fig:scalability}
	\vspace{-4ex}
\end{figure} 

\noindent\textbf{Collective Algorithm and Scalability.}
Figure~\ref{fig:scalability} illustrates the impact of algorithm selection and scalability on both \accl and software MPI during collective executions.
For an 8 KB message size, \accl's reduce operation adopts an all-to-one algorithm, resulting in minimal latency increase across nodes. However, recognizing potential bottlenecks at the root node with this approach, \accl switches to a binary tree algorithm for larger message sizes, such as 128 KB. In this case, an increase in latency is observed after four nodes, stabilizing until eight nodes due to a consistent tree depth.
On the other hand, software MPI exhibits a more fine-grained approach to algorithm selection based on the scale of the message size and the number of nodes. For instance, it deploys three distinct algorithms within the 8 KB range: an all-to-one algorithm for fewer than four nodes, a ring protocol for four to eight nodes, and an optimized binomial algorithm for 8 nodes. Additionally, for larger messages, software MPI switches between an all-to-one algorithm below three nodes and a binomial tree algorithm between four and eight nodes. This fine-grained algorithmic tuning contributes to its superior performance in certain H2H scenarios.
While software MPI's approach involves detailed algorithmic tuning, \accl's flexible design allows for potential future enhancements through additional fine-grained tuning to further optimize performance.

\begin{figure}[t]
	\centering
	\subfloat[Gather.]{\includegraphics[width=1.625in]{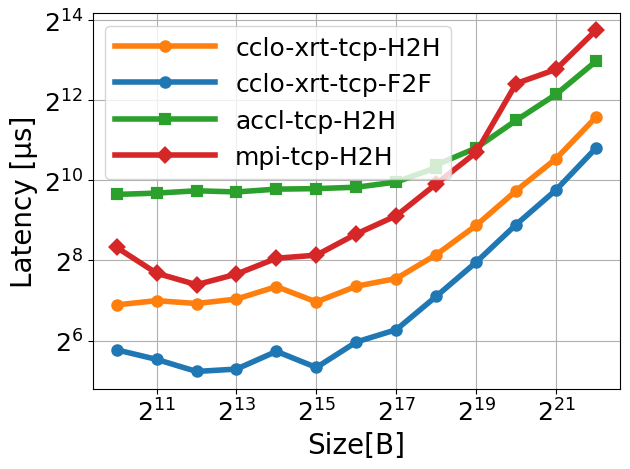} 
		\label{fig:tcp_comparison_gather}} 
	\subfloat[Reduce.]{\includegraphics[width=1.625in]{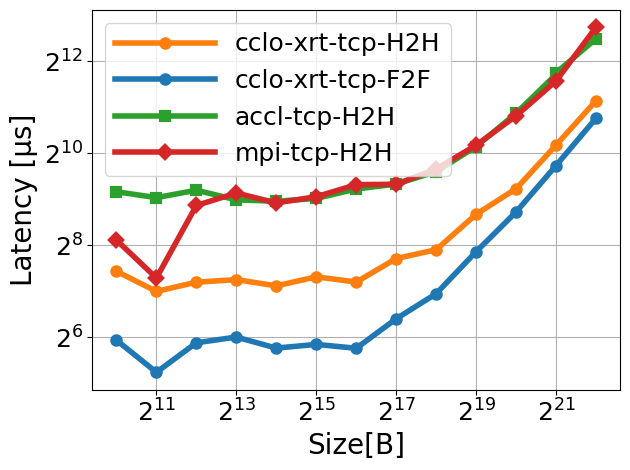} 
		\label{fig:tcp_comparison_reduce}} 
		\vspace{-1.5ex}
		\caption{Comparison of collective performance between \accl TCP with XRT, software MPI TCP and ACCL TCP.}
            \label{fig:tcp_comparison}
	\vspace{-2ex}
\end{figure}

\noindent\textbf{XRT Platform and TCP.}
In Figure~\ref{fig:tcp_comparison}, we evaluate \accl TCP with the XRT platform and compare it against software MPI with TCP. We also include a comparison with ACCL~\cite{ACCL} collectives, which employs a similar embedded processor to orchestrate collectives and supports TCP on the XRT platform. Notably, \accl TCP consistently outperforms its software counterpart across all configurations, benefiting from the line-rate processing capabilities of a hardware TCP POE. Furthermore, \accl demonstrates superior performance compared to ACCL. While both \accl and ACCL utilize embedded microprocessors for collective orchestration in hardware, \accl distinguishes itself by offloading more tasks to the hardware data plane, such as utilizing the Rx Buffer Manager for packet assembling. In contrast, ACCL relies more on the microprocessor, leading to lower performance. When comparing \accl TCP for serving host applications and device applications, a significant overhead is observed for host applications. This is attributed to the limitation of XRT platform, which prohibits direct data movement from the FPGA kernel to host buffers, resulting in a memory-copy overhead. Additionally, the XRT software invocation latency is notably higher, as indicated in Figure~\ref{fig:invocation_latency}.

\vspace{-1ex}
\section{Case Study: Deep Learning Recommendation Model}

\begin{table}[t]
    \centering
    \caption{Parameters of the target recommendation model.}\label{tab:parameters}
    \setlength{\tabcolsep}{1.0mm}{
    \begin{tabular}{ccccc}
    \toprule
        & Tables & Concat Vec Len & FC Layers & Embed Size\\
        & 100               & 3200                 & (2048, 512, 256)     & 50GB\\
    \bottomrule
    \end{tabular}}
    \vspace{-2ex}
\end{table}

Deep Learning Recommendation Models (DLRM) are widely used in personalized recommendation systems for various scenarios~\cite{gomez2015netflix, deep_youtube, din_alibaba_attention_fc}. 
The structure of a DLRM includes two major components: memory-bound embedding layers and computation-bound fully-connected (FC) layers. These models handle both dense and sparse features, with the latter stored as embedding vectors in tables. In inference, these vectors are accessed via indexes, resulting in multiple random memory accesses. The retrieved embedding vectors are then concatenated with dense features and passed through several FC layers to predict the click-through rate, incurring heavy computational loads due to vector-matrix multiplication. 

DLRM has been a focal point for acceleration on GPUs and FPGAs, given that CPU solutions are generally constrained by both random memory access and computation~\cite{tensordimm_kaist, facebook_benchmark, hsia2020cross}. GPU-based solutions~\cite{hsia2020cross, gupta2020deeprecsys, hwang2020centaur, fleetrec} mostly accelerate the computation-bound FC layers to gain high throughput. However, the large batch sizes required for efficient GPU computation, coupled with random memory access, often lead to increased latency (tens of milliseconds). FPGA-based techniques~\cite{jiang2020microrec,Hetero-Rec} overcome the embedding lookup bottleneck by distributing tables across memory banks and enabling parallel accesses, leveraging high-bandwidth-memory (HBM) and on-chip memory (BRAM/URAM). However, this approach is constrained by the requirement for embedding tables to fit within a single FPGA's memory (e.g., 16 GB HBM on AMD Alveo-U55C), limiting the size of embedding layer. Additionally, the finite computational resources on a single node pose restrictions on overall throughput for all FC layers.

\subsection{Distributed DLRM Inference}
We aim to demonstrate that \accl can facilitate distributing DLRM inference across FPGAs to accommodate larger embedding layers, as in many large-scale industrial settings, while at the same time achieving low latency and high throughput.
Table~\ref{tab:parameters} shows the detailed configuration of such an industrial-level recommendation model~\cite{fleetrec}. In such a use case, the embedding table does not fit into a single FPGA HBM and therefore both the embedding lookup and the computation are distributed across the network. This poses significant challenges for performance, scalability, and networking. 

\begin{figure}
    \centering
    \includegraphics[width=0.45\textwidth]{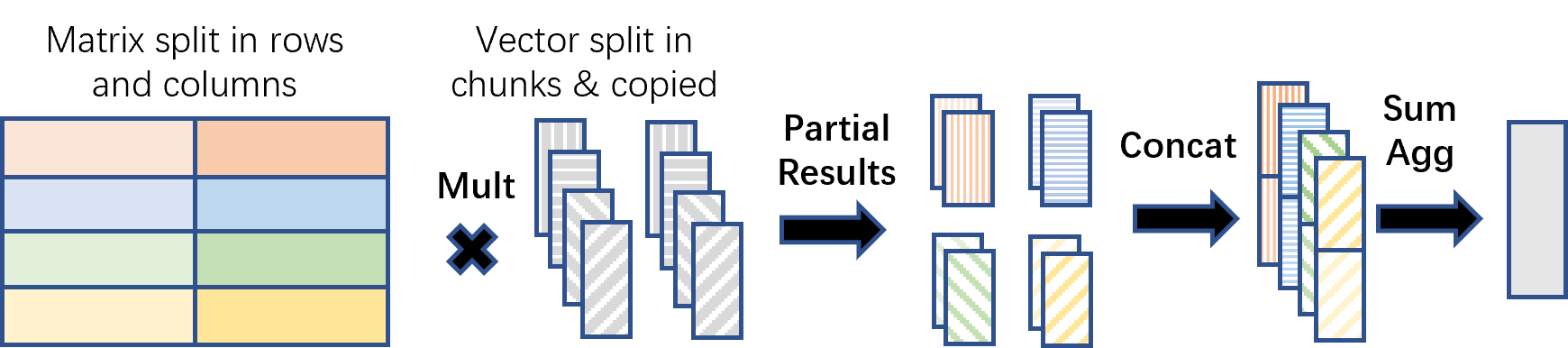}
    \caption{Checkerboard block decomposition.} \label{fig:matrix_decompose}
    \vspace{-2ex}
\end{figure}

\noindent\textbf{Vector-Matrix Multiplications Decomposition for DLRM.} 
The computation pattern in DLRM inference involves a chain of three vector-matrix multiplications, with the inference output vector computed as a sequence of operations involving three matrices of FC layers ($FC1$, $FC2$, $FC3$) and a concatenated embedding vector. The concept of distributed vector-matrix multiplication has been extensively studied in literature~\cite{vector-matrix} across CPUs and the same principle can be applied to an FPGA cluster. One common approach is \textit{checkerboard block decomposition of matrix}, as shown in Figure~\ref{fig:matrix_decompose}. This method involves partitioning the matrix in terms of both rows and columns, while partitioning the vector ensures that processes associated with the same matrix row partition share the same sub-embedding vector. Each process can then perform partial computations, and the results belonging to the same row partition are concatenated and subsequently aggregated. 

\begin{figure}
  \centering
    \includegraphics[width=.45\textwidth]{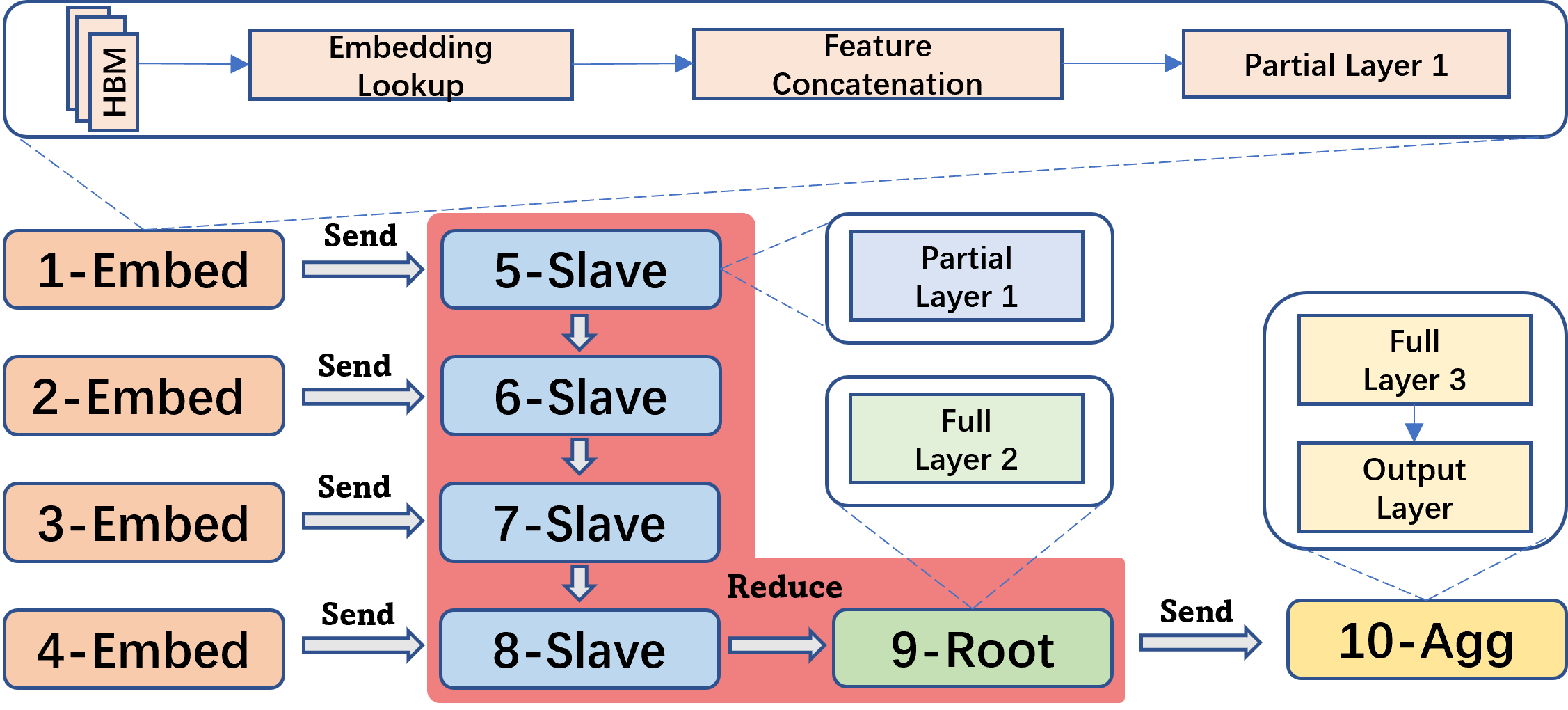} 
    \caption{Conceptual design of partitioned DLRM, with $FC1$ decomposed and $FC2$, $FC3$ pipelined across nodes.}
    \vspace{-2ex}
    \label{fig:dlrm_overview}
\end{figure}

\noindent\textbf{Decomposed and Pipelined Distributed DLRM.}
The partitioning strategy for the DLRM considers the need for balanced resource utilization, ensuring that the overall throughput is not limited by any process among all nodes. Typically, the computation load of the $FC1$ is significantly larger than subsequent layers like $FC2$ and $FC3$. To accommodate this, resource distribution should reflect the varying computation requirements. Additionally, for modern FPGAs with HBM, the capacity requires a minimum number of FPGAs to effectively store the embedding layer. A conceptual partitioned DLRM is illustrated in Figure~\ref{fig:dlrm_overview}. In this scenario, $FC1$ is decomposed and distributed across multiple FPGAs using the checkerboard block decomposition, and $FC2$ and $FC3$ are assigned to one FPGA each. The embedding tables are evenly distributed across nodes 1-4, with partial vectors transmitted to nodes 5-8, leveraging the network's low latency. Similarly, partial results computed on nodes 1-4 are forwarded to corresponding nodes 5-8, where an overall reduction of all partial $FC1$ results is conducted. The aggregated $FC1$ results are then forwarded to node 9 for $FC2$ computation, followed by node 10 for $FC3$ computation and final processing. Scaling resources according to the computation distribution requirements of each layer could lead to improved performance. For example, increasing the allocation of FPGAs for different layers based on their computational load.
Such partitioning method requires diverse communication patterns by each node, such as send-only, send/recv, and reduction and \accl provides a unified design supporting all the communication requirements of the DLRM through a standard interface. 
Additionally, for nodes that do not require reduction, the streaming reduction plugins of \accl can be removed with a compilation flag, reducing resource consumption and improving routing and timing. Furthermore, the cross-node simulation provided by \accl can facilitate the development process, reducing hardware debugging cycles.

\subsection{Use Case Evaluation}
\noindent\textbf{Distributed FC layer execution on CPU.} We use an illustrative example to demonstrate how \accl can be utilized to improve the efficiency of distributed work executing on CPU. In this use case, we distribute an FC layer workload (matrix-vector multiplication) by partitioning the weight matrix column-wise, with each rank receiving part of the input vector and a subset of the weight matrix columns. The matrix-vector product is obtained by summing the partial rank products using the reduce collective. For the implementation, we use the highly optimized Eigen library ~\cite{guennebaud2010eigen}, distributing it with both \accl RDMA and MPI RDMA. In this experiment we do not overlap computation and communication. 

\begin{figure}
  \centering
  \includegraphics[width=0.475\textwidth]{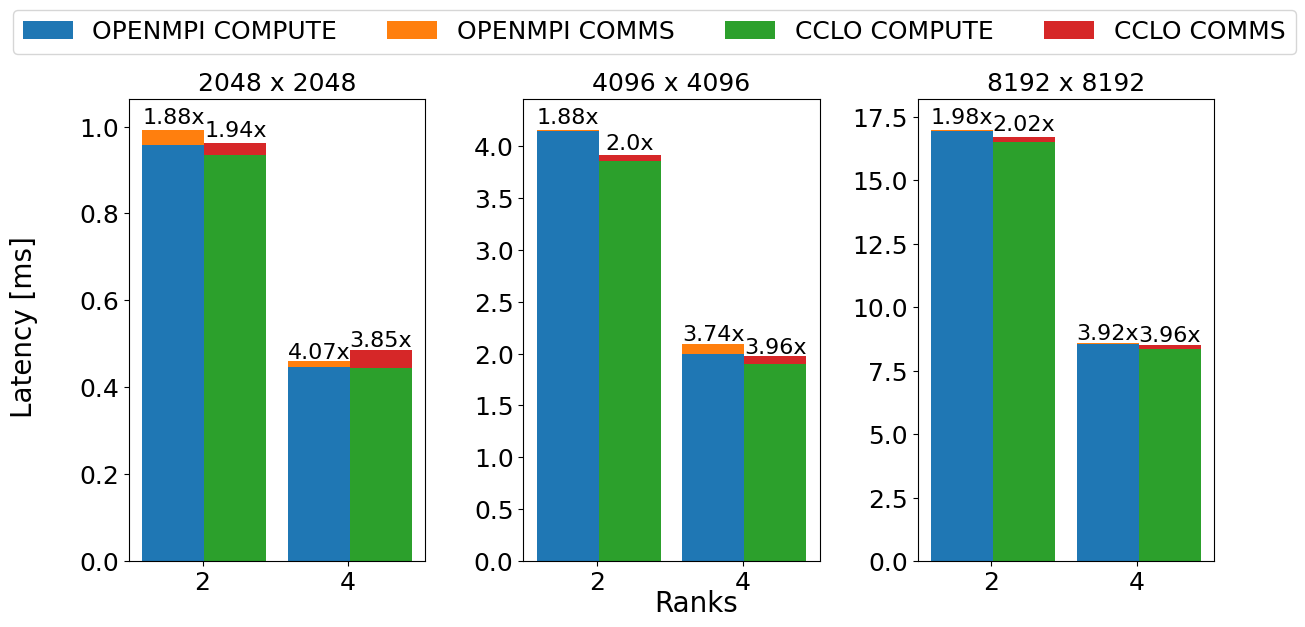}
  \caption{Speedup comparison and latency breakdown of distributed vector-matrix multiplication.}
  \vspace{-3ex}
  \label{fig:matrix-mult}
\end{figure}

The overall execution time of the distributed FC layer is compared to its single-node execution, as depicted in Figure~\ref{fig:matrix-mult}, where top-of-bar numbers indicate the speed-up compared to single-node execution. We observe that utilizing \accl instead of MPI for the reduction generally results in lower matrix-vector computation time. This performance increase is most likely due to reduced pressure on the CPU cache, as \accl utilizes FPGA memory for all intermediate reduction data structures. The figure indicates two instances of super-linear scaling, attributed to the weight matrix partitions fitting into either L2 (8 MB) or L3 (128 MB) caches on the CPU after partitioning, whereas the entire matrix did not fit in caches during single-node execution.
The reduction time itself is higher in most cases due to an additional copy required to move data between Eigen result buffers and \accl buffers, which can be eliminated with further optimization.
Overall, distributing work with \accl achieves lower latency, especially for specific configurations of FC size and number of ranks.

\begin{figure}[t]
	\centering
	\subfloat[DLRM latency.]{\includegraphics[width=1.625in]{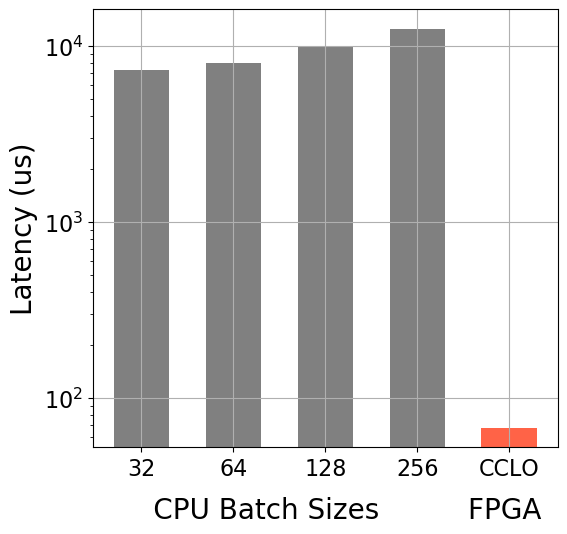} 
		\label{fig:latency}} 
	\subfloat[DLRM Throughput.]{\includegraphics[width=1.625in]{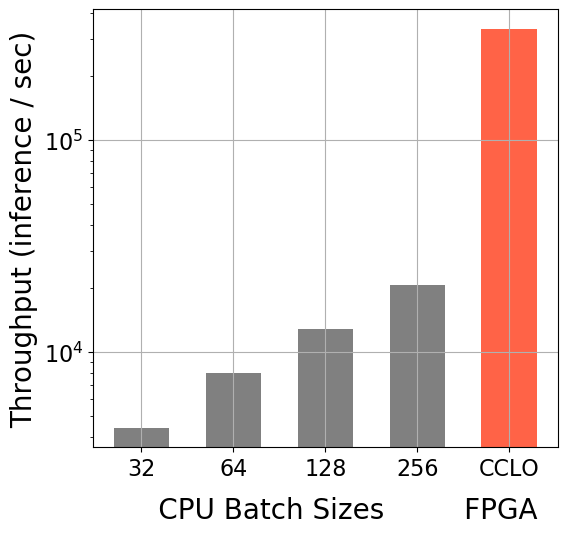} 
		\label{fig:throughput}} 
		\vspace{-1.5ex}
		\caption{\accl DLRM performance comparison.}
            \label{fig:dlrm}
	\vspace{-2ex}
\end{figure}

\noindent\textbf{Distributed FPGA-based DLRM.} We distribute an industrial DLRM model, as in Table~\ref{tab:parameters}, with \accl on 10 U55C FPGAs following the same design principle as shown in Figure~\ref{fig:dlrm_overview}. The communication between the embedding node and the reduce slave node during each inference requires the transmission of a 3.2 KB partial embedding vector and a 4 KB partial result. Additionally, the reduction process spanning nodes 5 to 9 operates with a message size of 8 KB per inference. The achieved operating frequency is 115 MHz. We utilize 32-bit fixed-point precision for computation. All the application kernels utilize streaming collective APIs to interact with \accl. \accl DLRM is configured with the TCP backend from XRT. Though the communication latency could be further optimized with \accl RDMA, it is not on the critical path of overall latency as it is overlapped with computation.
We also compare with CPU implementation~\cite{fleetrec}, where the DLRM inference is run on an Intel Xeon Platinum 8259CL CPU @ 2.50 GHz (32 vCPU, Cascade Lake, SIMD supported) and 256 GB DRAM with TensorFlow Serving enabled.
Figure~\ref{fig:dlrm}(a) shows the latency comparison between \accl and the CPU baseline. We evaluate various batch sizes on the CPU. On the other hand, \accl works with streaming data without batching. The hardware implementation demonstrates two orders of magnitude lower latency compared to the CPU. This substantial latency reduction in the hardware implementation is attributed to the parallel arithmetic units in hardware and the significant latency introduced by random memory accesses.
Figure~\ref{fig:dlrm}(b) shows the throughput comparison. \accl shows more than an order of magnitude higher throughput compared to CPU baseline. 

\subsection{Resource Consumption}
The resource utilization of \accl components and the overall utilization of DLRM across nodes are summarized in Table~\ref{tab:resources}. In the \accl subsystem, the majority of resources are allocated to POEs, with the TCP POE being the most resource-intensive, while the CCLO engine utilizes comparatively fewer LUT and memory resources.
DLRM utilization is categorized by different layers, and the presented utilization values represent the sum across multiple FPGAs after decomposition. Note that DLRM $FC1$ utilization exceeds 100\%, reflecting the decomposition across 8 FPGAs (max 800\%). The primary resource bottlenecks for DLRM are URAM, serving as fast on-chip memory for storing small embedding tables, and DSP, essential for matrix computations.

\begin{table}[]
\small
\begin{center}
\caption{Resource utilization.}
\label{tab:resources}
\begin{tabular}{ccccc}
\toprule
\textbf{Component} & \textbf{CLB kLUT} & \textbf{DSP} & \textbf{BRAM} & \textbf{URAM} \\ \midrule
\textbf{U55C(100\%)} & 1303 & 9024 & 2016 & 960\\ 
\midrule
\textbf{CCLO} & 12.1\% & 1.6\% & 5.7\% & 0 \\
\textbf{TCP POE} & 19.8\% & 0 & 10.6\% & 0 \\
\textbf{RDMA POE} & 13.0\% & 0 & 5.3\% & 0 \\
\midrule
\textbf{DLRM FC1} & 278.1\% & 580.1\% & 186.3\% & 798.3\% \\ 
\textbf{DLRM FC2} & 29.6\% & 85.1\% & 34.2\% & 97.9\% \\ 
\textbf{DLRM FC3} & 6.2\% & 16.1\% & 2.2\% & 20.8\% \\ 
\bottomrule
\end{tabular}
\end{center}
\vspace{-4ex}
\end{table}

\vspace{-1ex}
\section{Conclusion}
In this paper, we introduce \accl, an open-source FPGA-based collective library designed for portability across diverse platforms and communication protocols. \accl offers flexibility in implementing collectives without the need for FPGA re-synthesis and demonstrates high performance as collective abstractions for FPGA-distributed applications and as a collective offload engine for CPU applications. With \accl, there is potential for exploring new possibilities by extending collectives across CPU and FPGA boundaries and orchestrating them for a unified application.
\vspace{-1ex}

\clearpage
\bibliographystyle{plain}
\bibliography{references}

%%%%%%%%%%%%%%%%%%%%%%%%%%%%%%%%%%%%%%%%%%%%%%%%%%%%%%%%%%%%%%%%%%%%%%%%%%%%%%%%
\end{document}